\documentclass[preprint,aps,prd,groupedaddress,showpacs,nofootinbib]{revtex4-1}%
\usepackage{graphicx}
\usepackage{amsmath}
\usepackage{amssymb}
\usepackage{bm}
\usepackage{epsfig}
\usepackage{dcolumn}
\usepackage{slashed}
\usepackage{color}

\def\LO{\text{LO}}
\def\NLO{\text{NLO}}
\def\NNLO{\text{NNLO}}

\begin{document}

\def\Maryland{Maryland Center for Fundamental Physics, University of Maryland, College Park, Maryland 20742, USA}
\def\Argonne{High Energy Physics Division, Argonne National Laboratory, Argonne, IL 60439, USA}
\def\Northwestern{Department of Physics \& Astronomy, Northwestern University, Evanston, IL 60208, USA}

\title{$W$-boson plus jet differential distributions at NNLO in QCD}

\author{Radja Boughezal}
\email{rboughezal@anl.gov}
\affiliation{\Argonne}

\author{Xiaohui Liu}
\email{xhliu@umd.edu}
\affiliation{\Maryland}

\author{Frank Petriello}
\email{f-petriello@northwestern.edu}
\affiliation{\Argonne}
\affiliation{\Northwestern}

\begin{abstract}

We present a detailed phenomenological study of $W$-boson production in association with a jet through next-to-next-to-leading order (NNLO) in perturbative QCD.  Fiducial cross sections and differential distributions for both 8 TeV and 13 TeV LHC collisions are presented, as are results for both the inclusive one-jet bin and the exclusive one-jet bin.  Two different event selection criteria are considered: a general selection with standard cuts used in experimental analyses, and a boosted selection that focuses on high transverse momentum jets.  We discuss the higher-order corrections in detail and identify for which observables and phase space regions the QCD perturbative expansion is under good theoretical control, and where additional work is needed.  For most distributions and phase space regions the QCD perturbative expansion exhibits good convergence after the inclusion of the NNLO corrections.

\end{abstract}

\maketitle

\section{Introduction}

The production of $W$-bosons in association with jets plays a key role in the physics program of the Large Hadron Collider (LHC).  The accessible jet transverse momenta in this process extend beyond 1 TeV, allowing the predictions of the electroweak sector of the Standard Model to be tested in a previously inaccessible energy range.  Electroweak perturbative corrections become large in this region due to Sudakov logarithms that grow with energy.  Their interplay with perturbative QCD effects becomes important and interesting to explore.  In addition to these motivations, $W$-boson plus multi-jet production is a dominant background to signals for supersymmetry, while $W$-boson production with one or more jets serves as an important background to dark matter production in the mono-jet channel.  

Numerous theoretical studies of the $W$+jet process at higher orders in perturbation theory have been performed over the years.  The next-to-leading order (NLO) corrections in the strong coupling constant have been known for some time~\cite{Giele:1993dj}.  The NLO electroweak corrections were considered in Ref.~\cite{EWcor}.  The leading threshold logarithms beyond NLO in QCD have been considered~\cite{Becher:2011fc}.  Recently, a merged NLO QCD+electroweak prediction for $W$-boson in association with one, two or three jets was obtained~\cite{NLOQCDEW}.  Despite this significant progress, additional theoretical work is needed.  Large uncertainties still plague this process in certain kinematic regions.  At high transverse momenta, the contribution of di-jet events with the emission of soft and/or collinear gauge bosons leads to the appearance of NLO QCD corrections that reach a factor of 100 or more~\cite{Rubin:2010xp,Bauer:2009km}.  Although these large shifts can be removed by imposing a jet veto on events with two or more jets, such a veto leads to large logarithms that grow with energy~\cite{Boughezal:2015oga}.  The exclusive $W$+jet cross section is itself interesting to understand, as it is similar to exclusive jet-bin cross sections important in Higgs boson searches~\cite{Dittmar:1996ss}.  Methods for resumming jet-veto logarithms can therefore be validated using the $W$+jet process.  Measurement of the $W$+jet process at high transverse momentum aids in the development and validation of jet substructure techniques~\cite{boostcms,boostatlas}.  For all of these reasons, we must bring the perturbative series for $W$+jet under better theoretical control, in order for this process to serve as a precision benchmark for LHC studies.  

Recently, the full next-to-next-to-leading order (NNLO) QCD corrections to the $W$+jet process were obtained~\cite{Boughezal:2015dva}.  This calculation represents the first precision QCD prediction with a reliable theoretical error estimate for this process. Although various approximations reproduce the high-energy region where giant $K$-factors dominate, or capture certain classes of logarithms, it is only through exact NNLO calculations that percent-level predictions are possible.  Comparison with the exact NNLO also allows these approximations, and complementary approaches such as merged NLO predictions for $W$+jets, to be tested.  Ref.~\cite{Boughezal:2015dva} provided predictions for 8 TeV LHC collisions in the energy region up to a couple of hundred GeV.  The estimated theoretical errors for the observables considered were at the few-percent level, suggesting that the perturbative QCD corrections to $W$+jet production are under excellent  control.  It is our intent to fully investigate the NNLO QCD corrections to determine the behavior of the perturbative expansion in all interesting phase-space regions.  We perform a detailed study of numerous distributions for $W$+jet production through NNLO in perturbative QCD for both 8 TeV and 13 TeV LHC collisions, with a focus on the high transverse momentum region accessible for the first time at the LHC.  These predictions are needed for both ongoing analyses of 8 TeV Run I data and upcoming Run II studies.  Results are provided for both the inclusive and exclusive 1-jet bins.  We consider both a general selection criterion and a ``boosted" selection criterion that mimic the cuts used in previous ATLAS analyses~\cite{boostatlas,Aad:2014qxa}.  We describe these cuts in detail in the next section, but note here that they differ primarily in the transverse momentum cut imposed on the leading jet (30 GeV for the general selection, and 500 GeV for the boosted selection).  We summarize below our primary findings.
\begin{itemize}

\item The fiducial cross sections in the inclusive one-jet bin receive modest NNLO corrections, 3\% for the general selection and 15\% for the boosted selection.  The NNLO shift is within the NLO error estimate, and the residual NNLO scale dependence is at the few-percent level.  In the exclusive one-jet bin, the NNLO correction reduces the NLO result by a few percent. 

\item The $W$-boson rapidity and leading-jet pseudorapidity distributions receive corrections that have little kinematic dependence, in both the inclusive and exclusive 1-jet bins.  The NNLO shift is within the NLO error estimate and the remaining theoretical error from uncalculated QCD corrections is at the few-percent level.  

\item The transverse momenta of the $W$-boson and leading jet in the inclusive one-jet bin receive modest corrections that grow slightly with $p_T$, reaching a maximum of $+15\%$ for the tail of the $p_T^{J_1}$ distribution.  In the exclusive 1-jet bin the NLO distributions become smaller than the NLO result by up to an order of magniutes.  The NNLO corrections are much smaller, and remain near unity for all transverse momenta.

\item The NNLO corrections to the $H_T$ distribution in the inclusive 1-jet bin are large, reaching $+75\%$ in the TeV region.  We note that the NNLO corrections to the $H_T$ distribution are necessary to bring fixed-order QCD predictions into agreement with the 7 TeV data~\cite{Boughezal:2016yfp}.

\item QCD perturbation theory is under good control for the boosted selection after the inclusion of NNLO corrections, except near kinematic boundaries where soft gluon emission dominates.

\end{itemize}
We conclude that for most observables and phase space regions, the perturbative QCD expansion for $W$+jet is stabilized after the inclusion of the NNLO corrections; of course, the detailed quantitative results we find are dependent on the selection cuts imposed on the final state.  The next step in precision theoretical studies of the $W$+jet process should combine the electroweak corrections with the NNLO QCD results to facilitate comparisons with high-energy Run II data, where electroweak Sudakov effects become increasingly important.

Our paper is organized as follows.  In Section~\ref{sec:setup} we summarize the parameters used in our numerical results.  We present results for the general selection criteria in Section~\ref{sec:numgen}, for both 8 TeV and 13 TeV LHC collisions and for numerous distributions in both the inclusive and exclusive one-jet bins.  In Section~\ref{sec:numboost} we present our numerical results for the boosted selection.  We conclude and summarize our results in Section~\ref{sec:conc}.

\section{Setup}
\label{sec:setup}

We discuss here our calculational setup for $W$-boson production in association with a jet through NNLO in perturbative QCD.  We study collisions at both an 8 TeV LHC and a 13 TeV LHC, and consider both the inclusive $\geq 1$-jet bin and the exclusive 1-jet bin.  Jets are defined using the anti-$k_t$ algorithm~\cite{Cacciari:2008gp} with $R=0.4$.  We use CT14 parton distribution functions (PDFs)~\cite{Dulat:2015mca} at the appropriate order in perturbation theory: LO PDFs together with a LO partonic cross section, NLO PDFs with a NLO partonic cross section, and NNLO with a NNLO partonic cross section.  We choose the central scale 
\begin{equation}
\mu_0 = \sqrt{M_{l\nu}^2+\sum_i (p_T^{J_i})^2}
\end{equation}
for both the renormalization and factorization scales, where $M_{l\nu}$ is the invariant mass of the $W$-boson and the sum $i$ runs over all reconstructed jets.  This dynamical scale correctly captures the characteristic energy throughout the entire kinematic range studied here, which extends into the TeV region.  To estimate the theoretical uncertainty we vary the renormalization and factorization scales independently in the range $\mu_0/2 \leq \mu_{R,F} \leq 2 \mu_0$, subject to the restriction
\begin{equation}
1/2 \leq \mu_R / \mu_F \leq 2.  
\end{equation}
All numerical results presented include both $W^{+}$ and $W^{-}$ contributions.  The following electroweak parameters are used in our numerical results:
\begin{equation}
M_Z = 91.1876 \, {\rm GeV},\;\;\; M_W = 80.398 \, {\rm GeV},\;\;\; \Gamma_W = 2.105 \, {\rm GeV}, \;\;\; G_F = 1.16639 \times 10^{-5} \, {\rm GeV}^{-2}
\end{equation}
The $G_F$ electroweak input scheme, in which all other couplings are derived from those shown above, is used.

We consider two selection criteria in this paper: a ``general" selection requirement that matches what CMS has used in their studies of the $W$+jet process at 13 TeV~\cite{Khachatryan:2016fue}\footnote{We thank Emanuela Barberis for communication regarding the selection cuts used by CMS.}, and a ``boosted" selection criterion that focuses on the high transverse momentum region and is designed to study the separation between jets and electroweak objects at high energies~\cite{boostatlas}.  We impose the following fiducial cuts for the general selection:
\begin{equation}
\begin{split}
p_T^{J} &> 30 \, \text{GeV}, \;\;\; |\eta^J|<2.4, \\
p_T^{lep} &> 25 \, \text{GeV}, \;\;\; p_T^{miss} > 25 \, \text{GeV}, \\
|\eta^{lep}| &< 2.4, \;\;\; m_T > 50 \, \text{GeV}.
\end{split}
\label{eq:cuts1}
\end{equation}
Here, $m_T$ refers to the transverse mass formed from the lepton transverse momentum and the missing transverse momentum.  $p_T^{J_1}$ refers to the transverse momentum of the leading jet, on which we have imposed an additional cut.  We study the following distributions for this selection: $p_T^{J_1}$, $p_T^W$, $\eta^{J_1}$, $Y_W$, $H_T$, and $m_T$.  Here, $Y_W$ denotes the rapidity of the $W$-boson, $\eta^{J_1}$ the pseudorapidity of the leading jet, and $p_T^{J_1}$, $p_T^W$ the transverse momenta of the leading jet and the $W$-boson, respectively.  $H_T$ is the scalar sum of the transverse momenta of all reconstructed jets.  The transverse mass $m_T$ is defined as
\begin{equation}
m_T = \sqrt{2p_T^{lep} p_T^{neut} \left( 1-\text{cos}(\phi_{l\nu})\right)},
\end{equation}
where $\phi_{l\nu}$ is the angle between the lepton and the neutrino in the transverse plane.  All of these distributions begin first at leading order for the $W$+jet process, and therefore the results presented here are genuine NNLO predictions.

For the boosted selection we impose the acceptance cuts:
\begin{equation}
\begin{split}
p_T^{J} &> 100 \, \text{GeV}, \;\;\; |\eta^J|<2.1,\;\;\; p_T^{J_1}>500 \, \text{GeV}, \\
p_T^{lep} &> 25 \, \text{GeV}, \;\;\; |\eta^{lep}| < 2.5.
\end{split}
\label{eq:cuts2}
\end{equation}
The primary change is that the transverse momentum cut on the leading jet has been increased to 500 GeV.  For this selection we will study the separation between the $W$-boson and the closest jet, and between the lepton and the closest jet.  The distance is measured using
\begin{equation}
\Delta R_{xy} = \sqrt{(\eta_x - \eta_y)^2+(\phi_x-\phi_y)^2},
\label{eq:delr}
\end{equation}
where $\phi_x$ denotes the transverse-plane azimuthal angle of particle $x$. 

The NNLO calculation upon which our phenomenological study is based was obtained using the $N$-jettiness subtraction scheme~\cite{Boughezal:2015dva,Gaunt:2015pea}.  This technique relies upon splitting the phase space for the real emission corrections according to the $N$-jettiness event shape variable, $\tau_N$~\cite{Stewart:2010tn}, and relies heavily upon the theoretical machinery of soft-collinear effective theory~\cite{scet}.  For values of $N$-jettiness greater than some cut, $\tau_N > \tau_N^{cut}$, an NLO calculation for $W$+2-jets is used.  Any existing NLO program can be used to obtain these results.  We use the public code MCFM~\cite{Campbell:2010ff,Campbell:2015qma} in this study.  For the phase-space region $\tau_N < \tau_N^{cut}$, an all-orders resummation formula is used to obtain the contribution to the cross section~\cite{Stewart:2010tn,Stewart:2009yx}.  An important check of the formalism is the independence of the full result from $\tau_N^{cut}$.  By now the application and validation of $N$-jettiness subtraction for one-jet processes has been discussed several times in the literature~\cite{Boughezal:2015dva,Boughezal:2015aha,Boughezal:2015ded}, and we do not review this topic here.  We note that we have computed each bin of the studied distributions for several $\tau_N^{cut}$ values, and have found independence of all results from $\tau_N^{cut}$ within numerical errors.

\section{Numerical results for the general selection}
\label{sec:numgen}

We begin by discussing the fiducial cross sections for both 8 TeV and 13 TeV collisions, assuming the general selection cuts of Eq.~(\ref{eq:cuts1}).  The LO, NLO, and NNLO inclusive 1-jet cross sections, as well as the $K$-factors $K_{\NLO} = \sigma_{\NLO}/\sigma_{\LO}$ and $K_{\NNLO} = \sigma_{\NNLO}/\sigma_{\NLO}$, are presented in Table~\ref{tab:fidinc}.  For both energies there is an approximately 40\% increase of the cross section in going from LO to NLO.  We will later see that a significant fraction of this increase occurs for high $H_T$, and arises from the contribution of dijet events that first occur at NLO.  The NNLO corrections are more mild, and increase the NLO result by only 3\% for the central scale choice.  This indicates the good convergence of QCD perturbation theory for the fiducial cross section.  The residual errors as estimated by scale variation decrease from the approximately 5\% level at NLO to the percent level at NNLO.  We note that the corrections here are slightly larger than those found in Ref.~\cite{Boughezal:2015dva}, which is due to the fixed scale choice used there.
 
\begin{table}[h]
\begin{tabular}{|c|c|c|c|c|c|}
\hline
 & $\sigma_{\LO}$ (pb) & $\sigma_{\NLO}$ (pb) & $\sigma_{\NNLO}$ (pb) & $K_{\NLO}$ & $K_{\NNLO}$ \\
 \hline\hline
8 TeV & $428.9^{+33.3}_{-31.5}$ & $611.1^{+38.1}_{-31.0}$ & $630.2^{+1.7}_{-6.8}$ & 1.42 & 1.03 \\
13 TeV & $773.7^{+33.7}_{-36.8}$ & $1099.3^{+57.8}_{-44.6}$ & $1130.2^{+5.2}_{-8.7}$ & 1.42 & 1.03 \\
\hline
\end{tabular}
\caption{Fiducial cross sections for the inclusive 1-jet bin for 8 TeV and 13 TeV collisions, using the cuts of Eq.~(\ref{eq:cuts1}).  The scale errors are shown for the LO, NLO and NNLO cross sections.  The $K$-factors are shown for the central scale choice.}
\label{tab:fidinc}
\end{table}

The fiducial cross sections in the exclusive 1-jet bin are shown in Table~\ref{tab:fidexc}.  The pattern of perturbative corrections for the exclusive 1-jet bin is different than the corrections seen in the inclusive 1-jet bin.  The NLO correction increases the LO result by 19\% for 8 TeV collisions, and by 16\% for 13 TeV collisions.  Including the NNLO terms decreases the cross section by about 3\% for both collision energies.  The origin of these different corrections are jet-veto logarithms, which are known to have a significant effect on fixed-order cross sections in exclusive jet bins~\cite{Stewart:2011cf,Banfi:2012yh}.  The relevant logarithm for this process is $\text{ln}(\sqrt{\hat{s}}/30 \, \text{GeV})$, where $\sqrt{\hat{s}}$ is the partonic center-of-mass energy and 30 GeV denotes the transverse momentum veto imposed on the additional jets.  Our cuts impose the minimum requirement
\begin{equation}
\sqrt{\hat{s}} \geq \sqrt{(p_T^{J_1,min})^2+M_{l\nu}^2}+p_T^{J_1,min} \approx 115 \, \text{GeV},
\label{eq:smin}
\end{equation}
where $p_T^{J_1,min} = 30 \, \text{GeV}$, indicating that these logarithms are sizable.    Although significant, the convergence of the $K$-factor when including NNLO corrections indicates that the effect of these logarithms on the fiducial cross section are still under control in fixed-order perturbation theory.  It would, however, be interesting to compare predictions obtained using the resummation formalism for the exclusive one-jet bin~\cite{Liu:2012sz,Liu:2013hba,Boughezal:2013oha,Boughezal:2015oga} with the NNLO results obtained here.

\begin{table}[h]
\begin{tabular}{|c|c|c|c|c|c|}
\hline
 & $\sigma_{\LO}$ (pb) & $\sigma_{\NLO}$ (pb) & $\sigma_{\NNLO}$ (pb) & $K_{\NLO}$ & $K_{\NNLO}$ \\
 \hline\hline
8 TeV & $428.9^{+33.3}_{-31.5}$ & $509.4^{+12.9}_{-12.0}$ & $495.9^{+3.5}_{-8.0}$ & 1.19 & 0.97 \\
13 TeV & $773.7^{+33.7}_{-36.8}$ & $895.7^{+16.0}_{-11.6}$ & $863.2^{+10.5}_{-13.0}$ & 1.16 & 0.96  \\
\hline
\end{tabular}
\caption{Fiducial cross sections for the exclusive 1-jet bin for 8 TeV and 13 TeV collisions, using the cuts of Eq.~(\ref{eq:cuts1}) and additionally imposed a veto on a second reconstructed jet.}
\label{tab:fidexc}
\end{table}

\begin{figure}[h]
    \includegraphics[width=.49\linewidth]{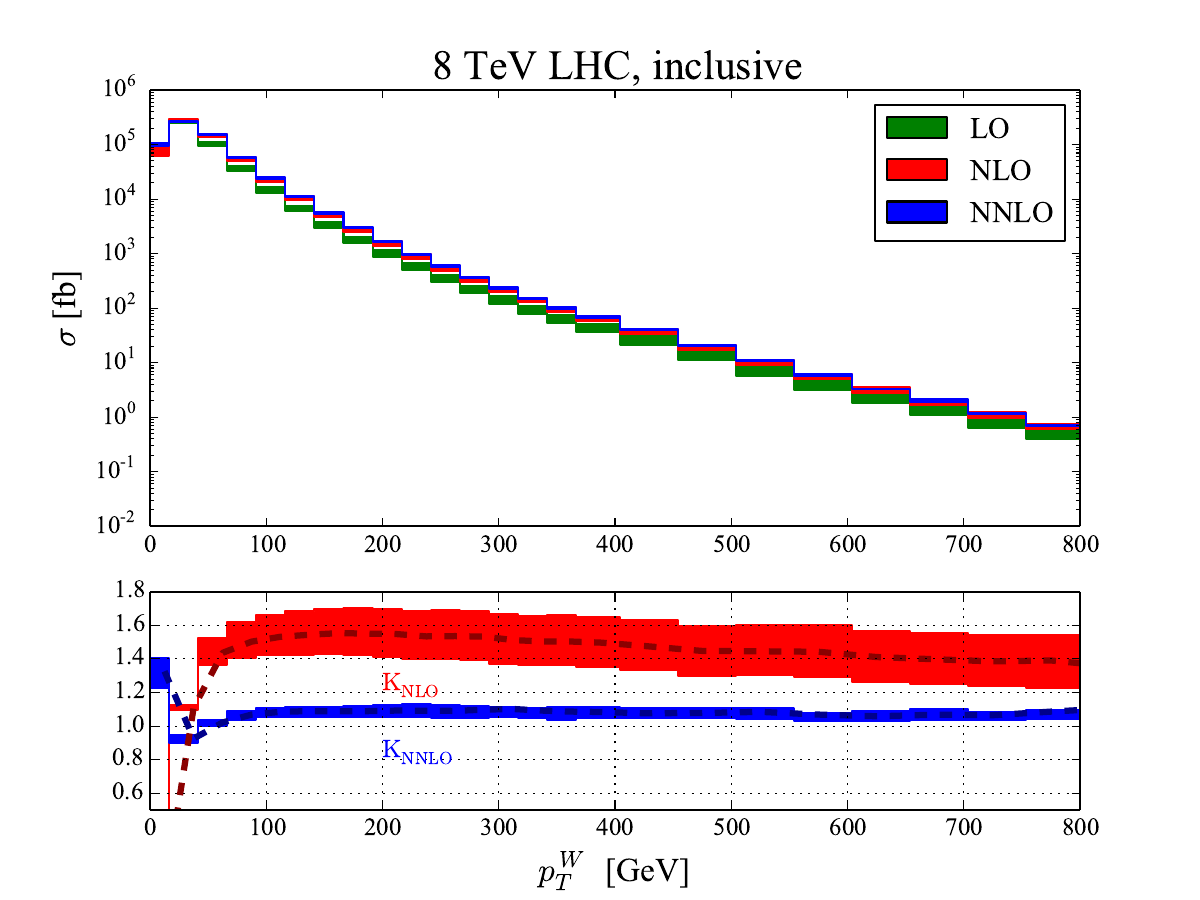} 
    \includegraphics[width=.49\linewidth]{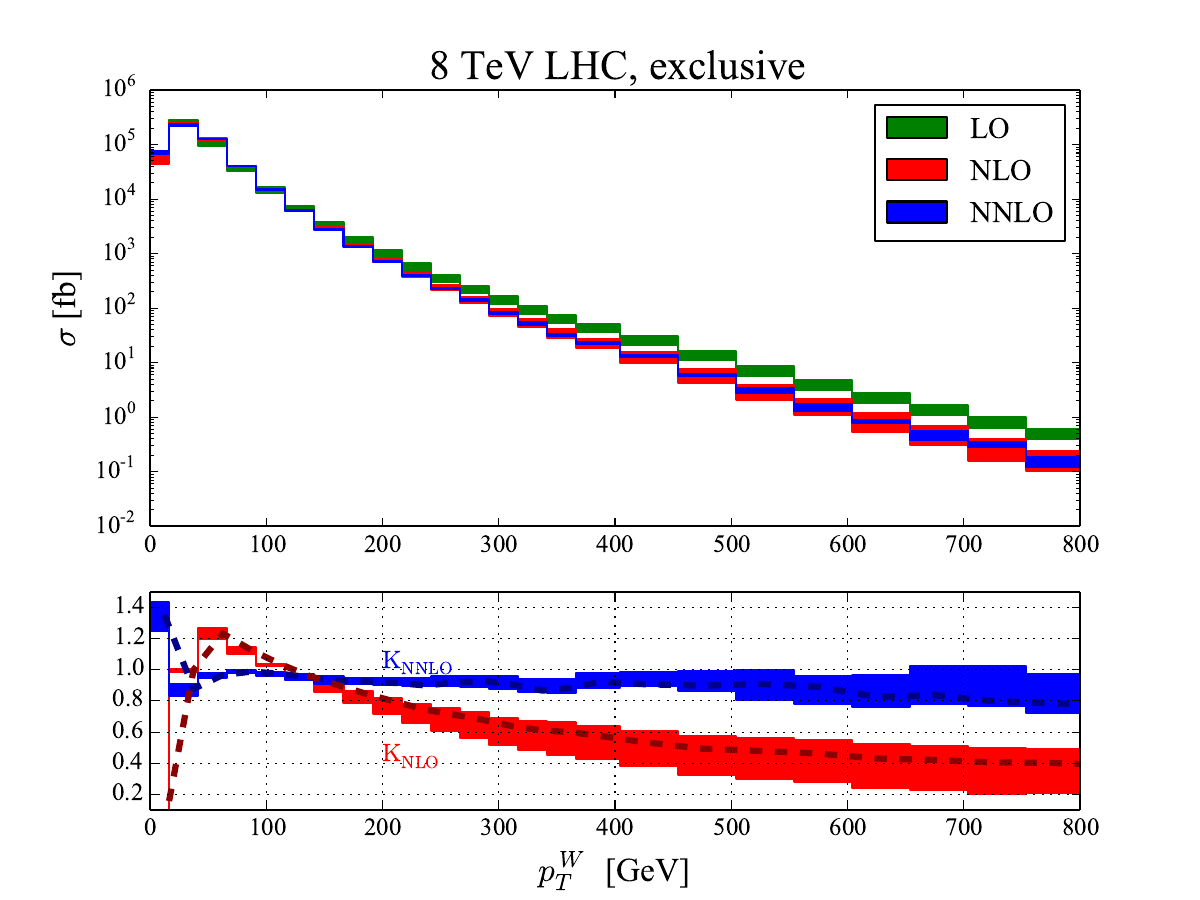} 
    \includegraphics[width=.49\linewidth]{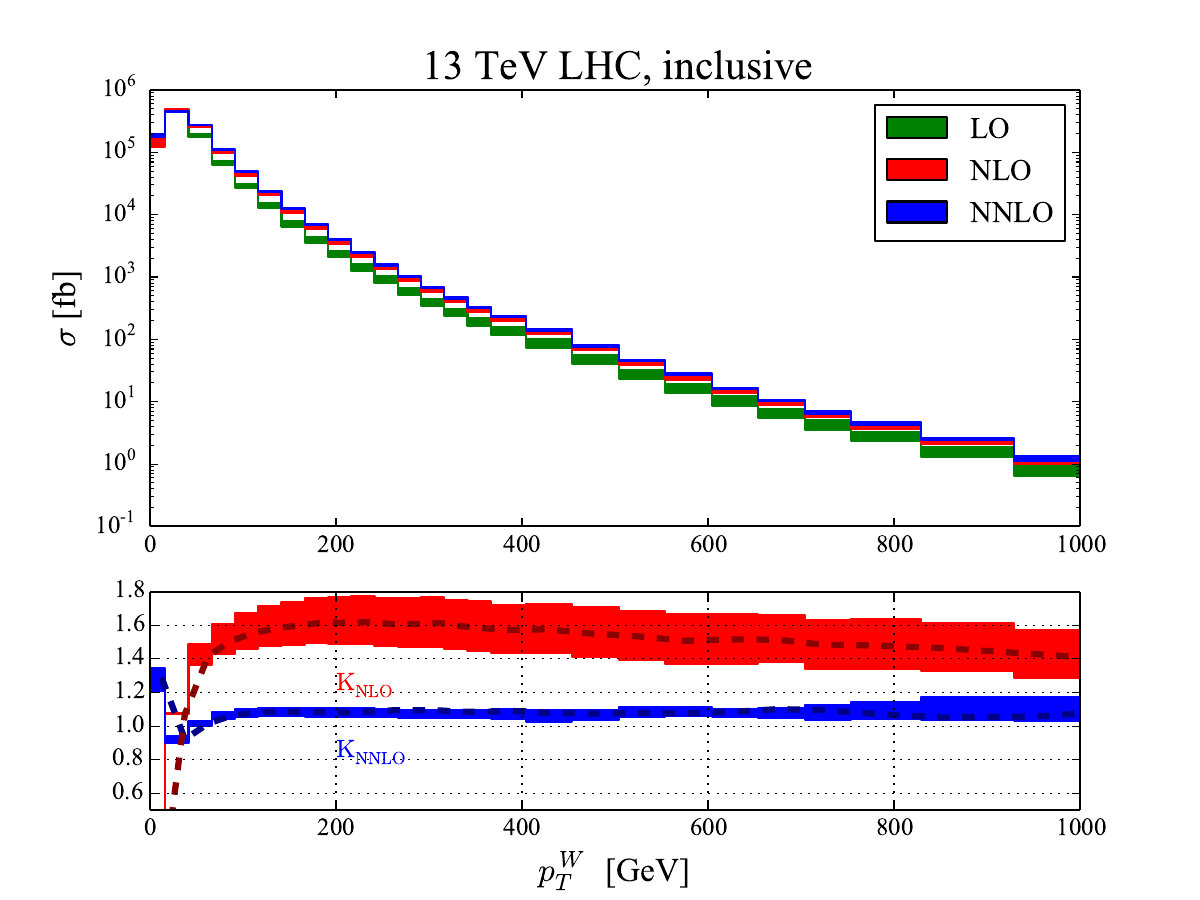} 
    \includegraphics[width=.49\linewidth]{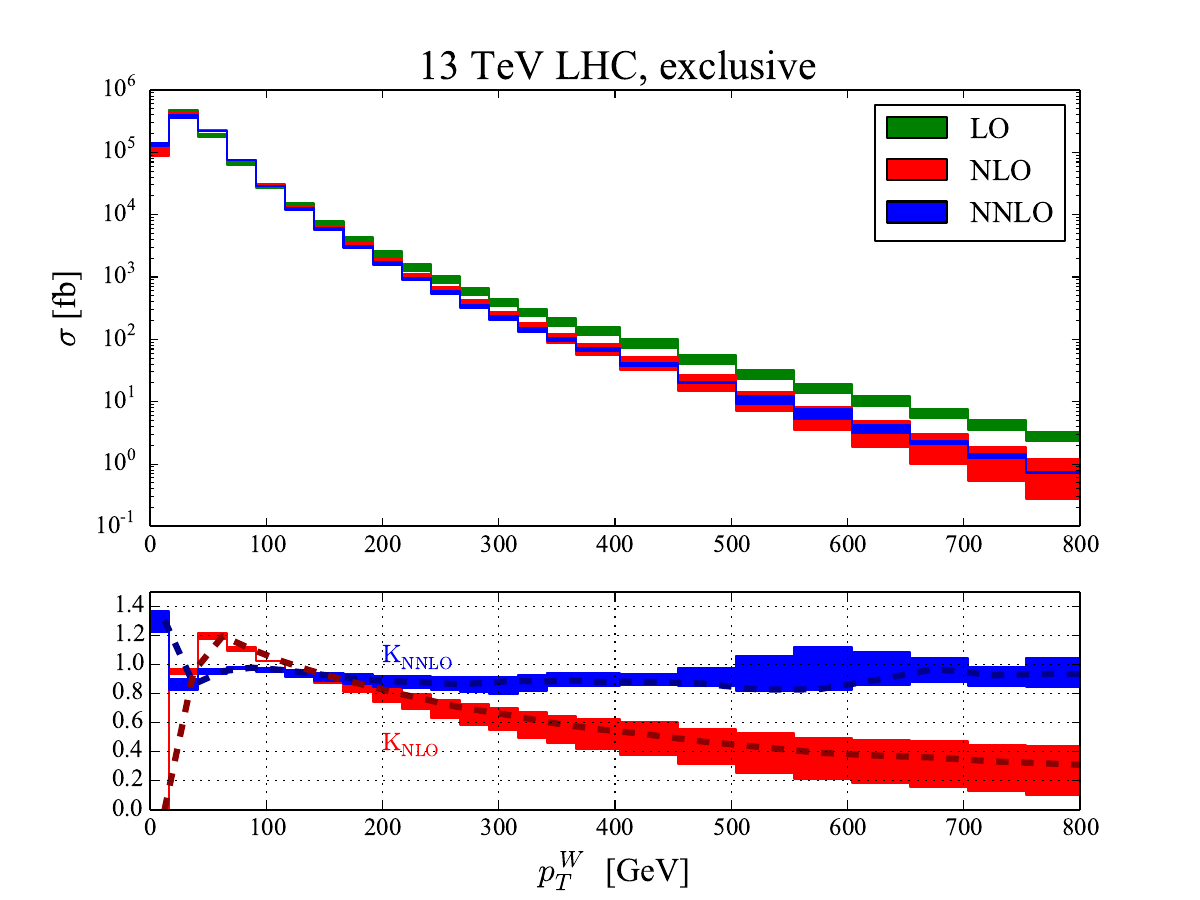} 
    \caption{Plots of the $W$-boson transverse momentum distribution for the following scenarios: 8 TeV inclusive 1-jet bin (upper left), 8 TeV exclusive 1-jet bin (upper right), 13 TeV inclusive 1-jet bin (lower left), 13 TeV exclusive 1-jet bin (lower right).  In each plot the upper inset shows the LO, NLO and NNLO distributions, while the lower inset shows $K_{\NLO}$ and $K_{\NNLO}$.  The bands indicate the scale variation, while the dashed lines in the lower panel indicate the result for the central scale choice.}
    \label{fig:pTW}
\end{figure}

We now study several distributions for both 8 TeV and 13 TeV collisions.   We begin with the $W$-boson transverse momentum distribution in Fig.~\ref{fig:pTW}.  Shown are the LO, NLO and NNLO distributions, as well as the associated $K$-factors, for both the inclusive and exclusive 1-jet bins.  The behavior of the inclusive $p_{TW}$ distribution is similar at both 8 TeV and 13 TeV.  For both 8 and 13 TeV collisions the NLO corrections increase the LO result by a maximum of roughly 60\% for the central scale choice at $p_{TW} \approx 200$ GeV, with the NLO shift decreasing to 40\% for $p_{TW} \approx 1$ TeV.  In both cases the NNLO correction increases to nearly 10\% for the central scale choice above $p_{TW}=200$ GeV, and remains approximately constant above this value.  We note that the leading-jet transverse momentum restriction $p_T^{J_1}> 30$ GeV implies that at LO, $p_{TW}>30$ GeV.  This restriction is relaxed at NLO.  Near this kinematic boundary the cross section is sensitive to soft-gluon radiation, leading to the observed large corrections for $p_{TW} \approx 30$ GeV.  The scale uncertainty above $p_{TW}=200$ GeV is approximately $\pm 20\%$ at NLO for both 8 TeV and 13 TeV. The NNLO estimated error becomes approximately $\pm 2-3\%$ above $p_{TW} = 200$ GeV for both 8 TeV and 13 TeV.
 
The exclusive 1-jet bin behaves very differently as a function of $p_{TW}$.  The NLO $K$-factor for 8 TeV collisions decreases from $K_{\NLO}\approx 0.9$ for $p_{TW}=150$ GeV to 0.4 at $p_{TW}=750$ GeV.  The NLO $K$-factor for 13 TeV collisions falls below 0.2 at high $p_T^W$.  As discussed around Eq.~(\ref{eq:smin}), the jet-veto logarithms increase with the transverse momentum, leading to the observed shape of the corrections. $K_{\NNLO}$ remains near unity for $p_T^W$ away from the kinematic boundary at 30 GeV.  It would be interesting to compare the fixed-order predictions with those of resummation-improved perturbation theory~\cite{Liu:2012sz,Boughezal:2015oga}.  
 
\begin{figure}[h]
    \includegraphics[width=.49\linewidth]{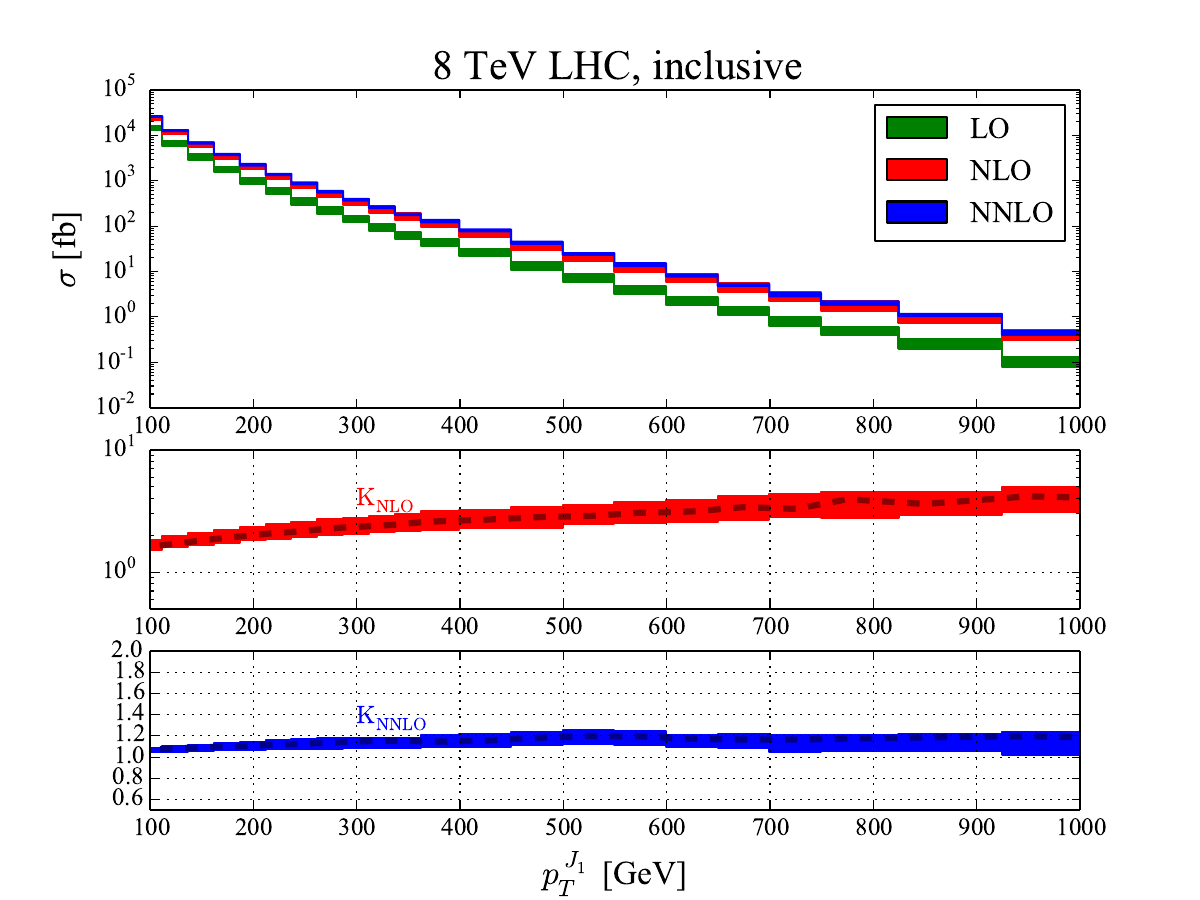} 
    \includegraphics[width=.49\linewidth]{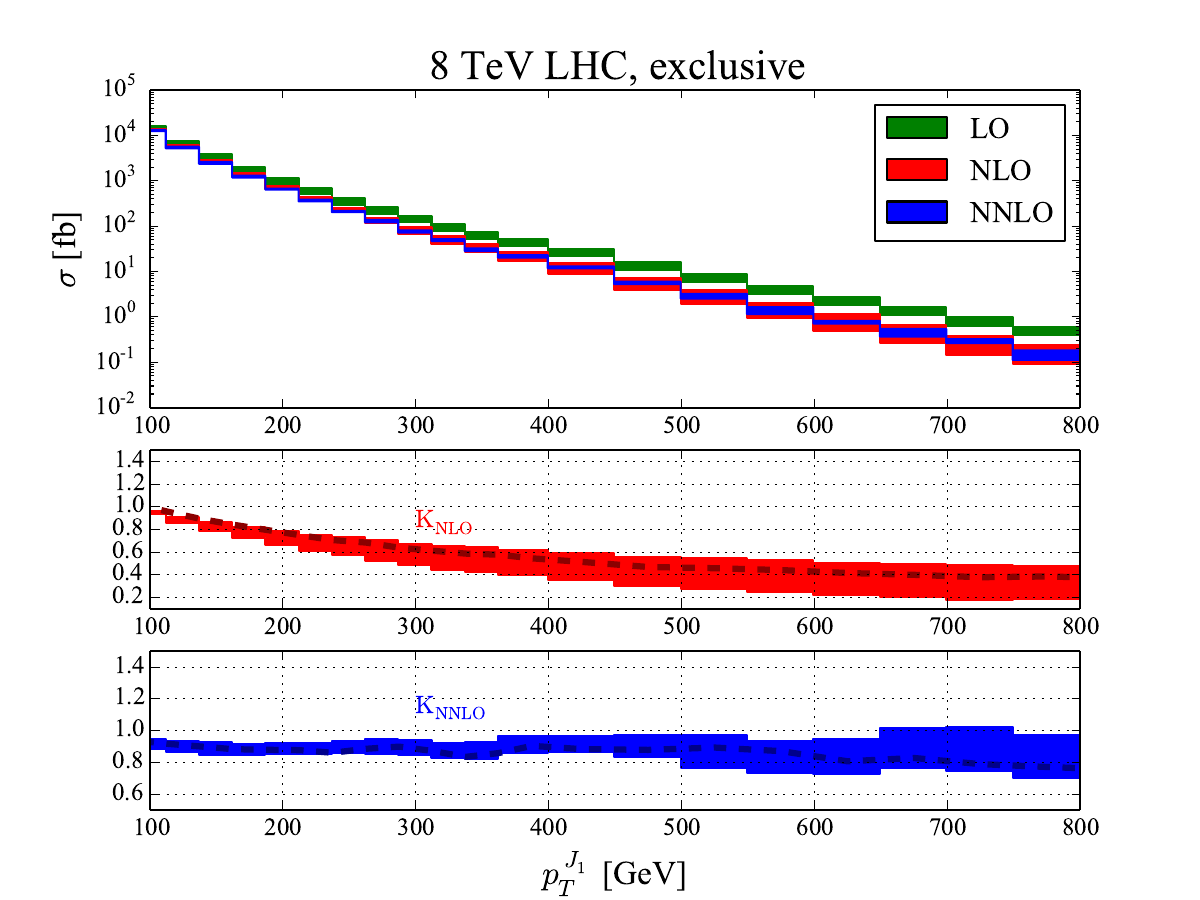} 
    \includegraphics[width=.49\linewidth]{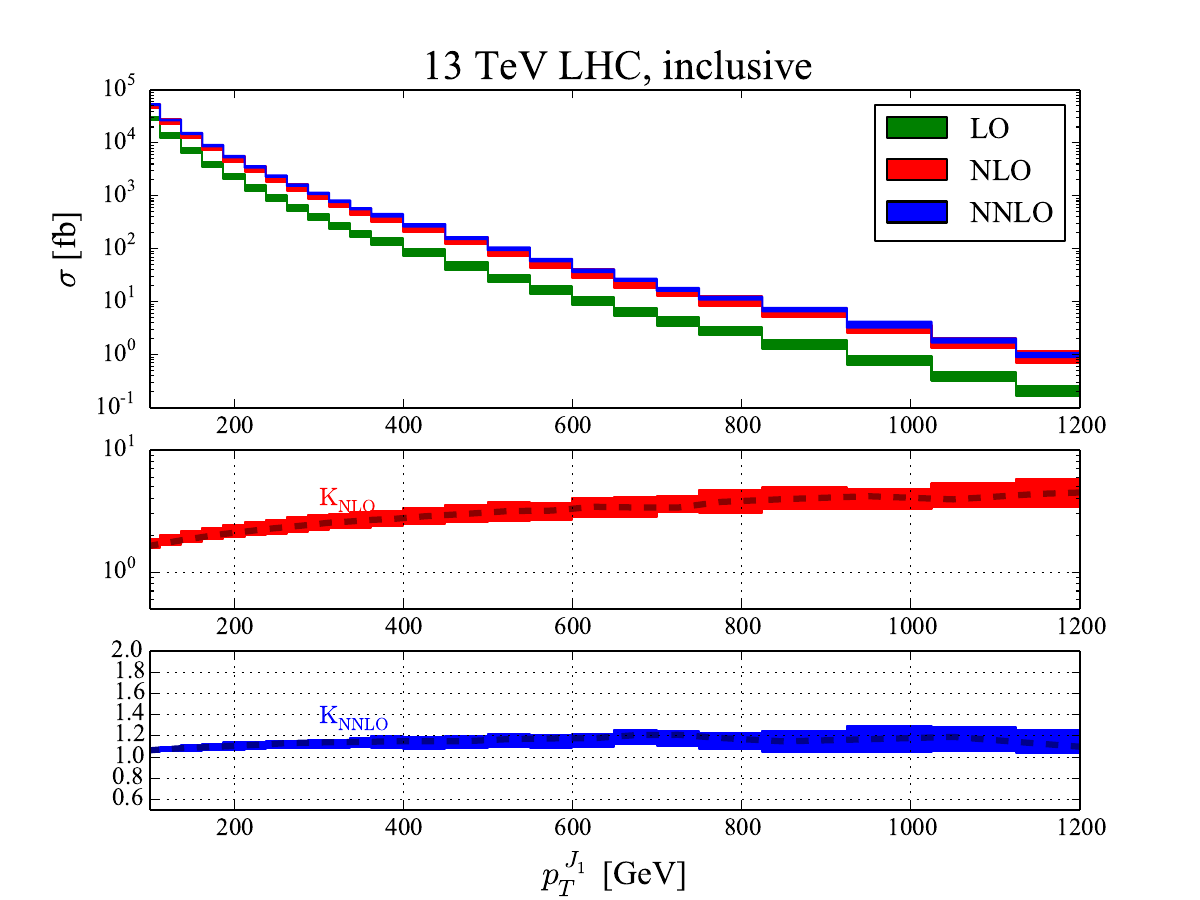} 
    \includegraphics[width=.49\linewidth]{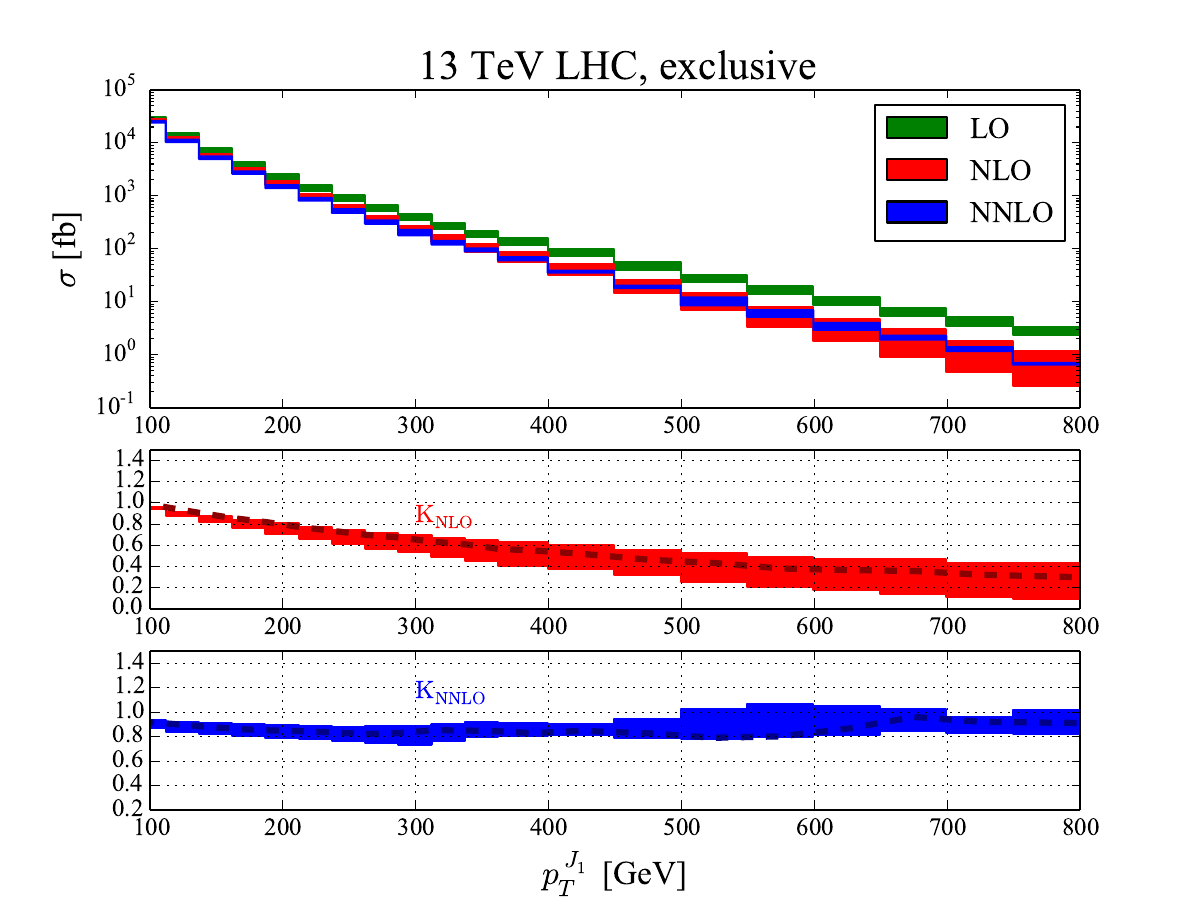} 
    \caption{Plots of the transverse momentum distribution of the leading jet for the following scenarios: 8 TeV inclusive 1-jet bin (upper left), 8 TeV exclusive 1-jet bin (upper right), 13 TeV inclusive 1-jet bin (lower left), 13 TeV exclusive 1-jet bin (lower right).  In each plot the upper inset shows the LO, NLO and NNLO distributions, while the lower inset shows $K_{\NLO}$ and $K_{\NNLO}$.  The bands indicate the scale variation, while the dashed lines in the lower panel indicate the result for the central scale choice.}
    \label{fig:pTJ1}
\end{figure}  
 
The transverse momentum distribution of the leading jet is presented in Fig.~\ref{fig:pTJ1}.  Shown are the LO, NLO and NNLO distributions, as well as the associated $K$-factors, for both the inclusive and exclusive 1-jet bins.  The first thing to note is the growth of the NLO $K$-factor with jet $p_T$.  It grows above a factor of four for $p_T^{J_1} > 1$ TeV for both 8 TeV and 13 TeV collisions.  The reason for these large corrections has been discussed in the literature~\cite{Rubin:2010xp,Bauer:2009km}.  At NLO there are configurations containing two hard jets and a soft/collinear $W$ boson that are logarithmically enhanced. These cannot occur at LO, since the $W$-boson must balance in the transverse plane against the single jet that appears.  Although the NLO corrections are large, the QCD perturbative expansion stabilizes when the NNLO corrections are included.  The additional increase at NNLO is more mild, rising from $+5\%$ for $p_T^{J_1}=100$ GeV to $+15\%$ for $p_T^{J_1}>500$ GeV, for both collision energies.  The NLO scale uncertainty is approximately $\pm 20\%$ in the several hundred GeV range of transverse momenta.  This decreases to $\pm 5\%$ at NNLO in the several hundred GeV range.  For $p_T^{J_1}<300$ GeV the NNLO uncertainty is at the few-percent level.  The pattern of corrections is similar for both 8 TeV and 13 TeV collisions.
 
The $p_T^{J_1}$ distribution in the exclusive 1-jet bin behaves similarly to the $p_T^W$ distribution.  For 8 TeV collisions the NLO $K$-factor falls to 0.4 at $p_T^{J_1}=700$ GeV, while it falls below 0.2 for 13 TeV collisions.  $K_{\NNLO}$ for 8 TeV collisions remains between 0.8 and 0.9 for the $p_T^{J_1}$ range studied. The scale dependence is at the 5\% level for $p_T^{J_1}<300$ GeV, and it gradually grows for higher transverse momenta.  We caution that fixed-order scale dependence may not be a good measure of the theoretical uncertainty in phase space regions containing large jet-veto logarithms~\cite{Stewart:2011cf}.
 
\begin{figure}[h]
    \includegraphics[width=.49\linewidth]{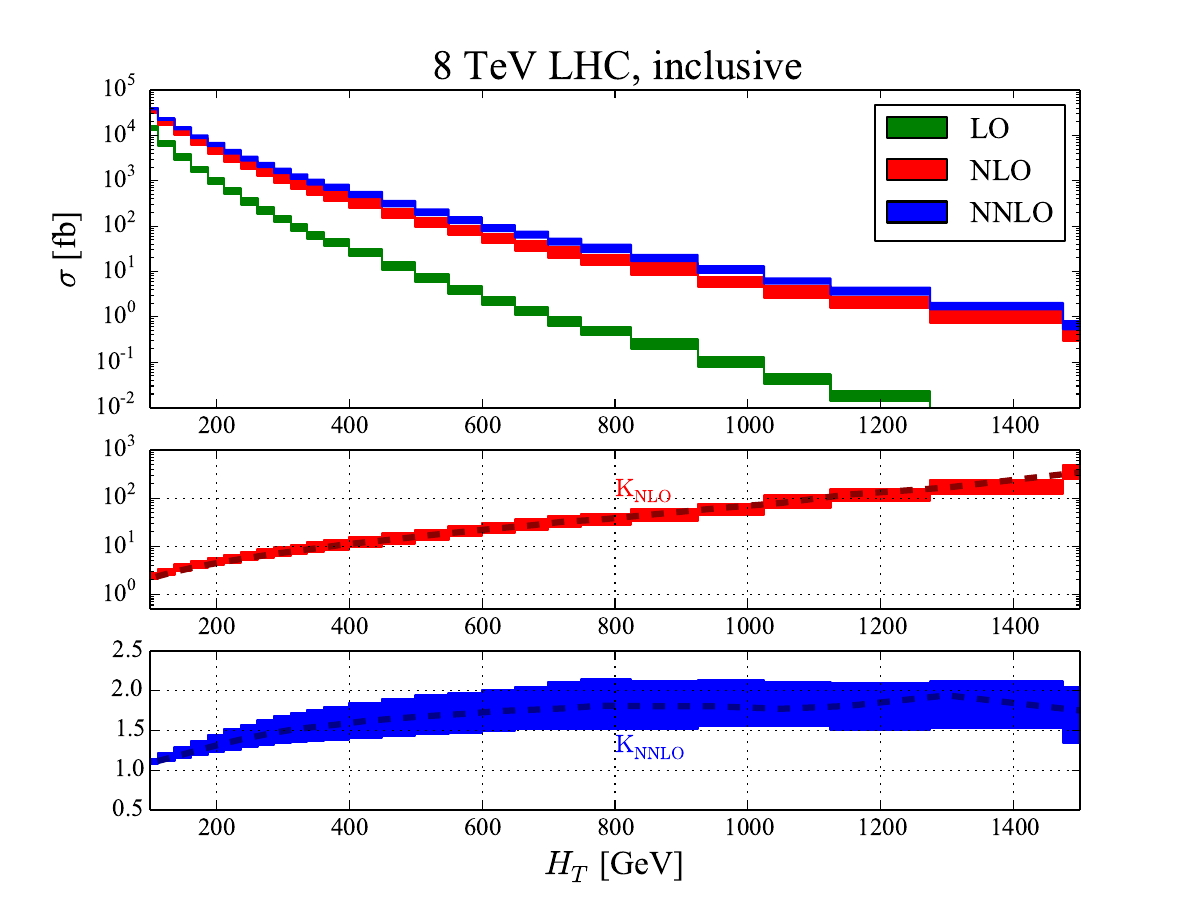} 
    \includegraphics[width=.49\linewidth]{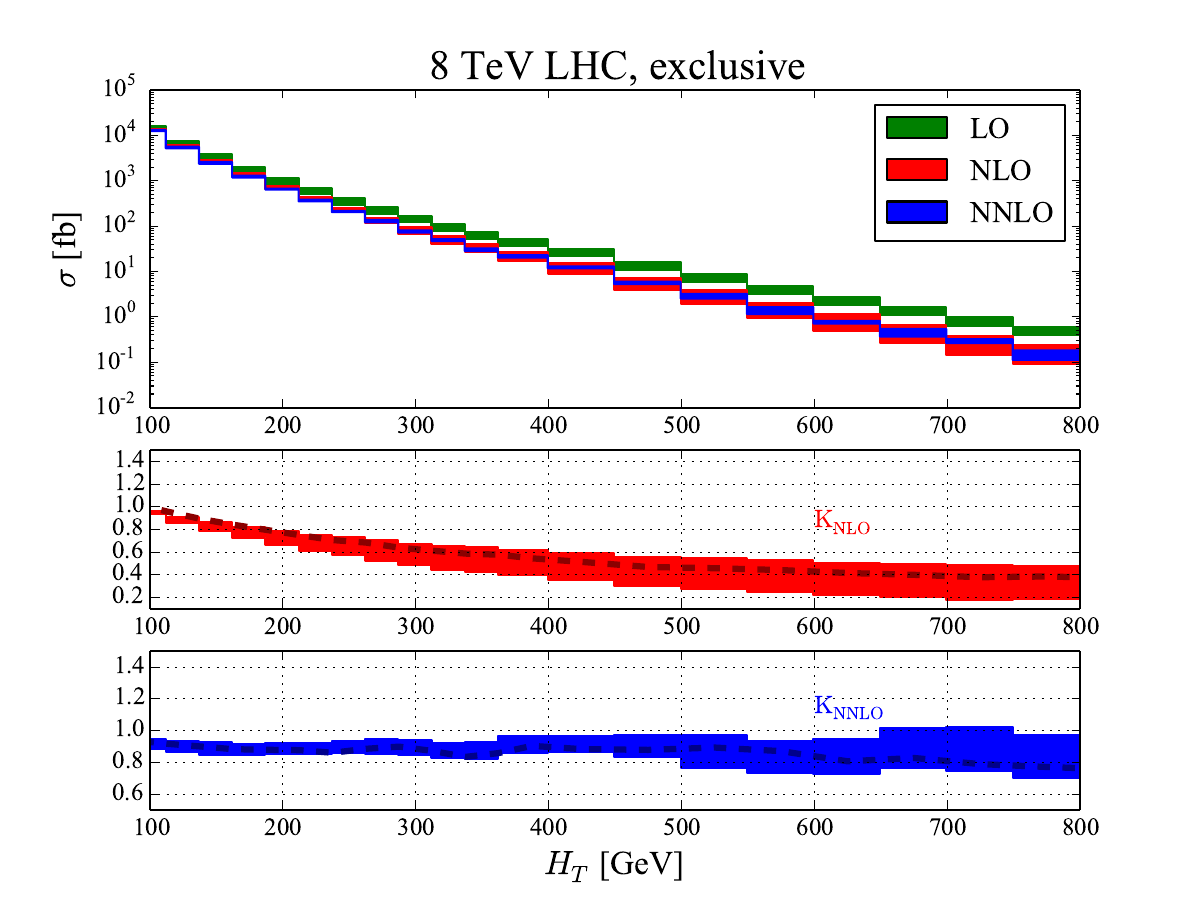} 
    \includegraphics[width=.49\linewidth]{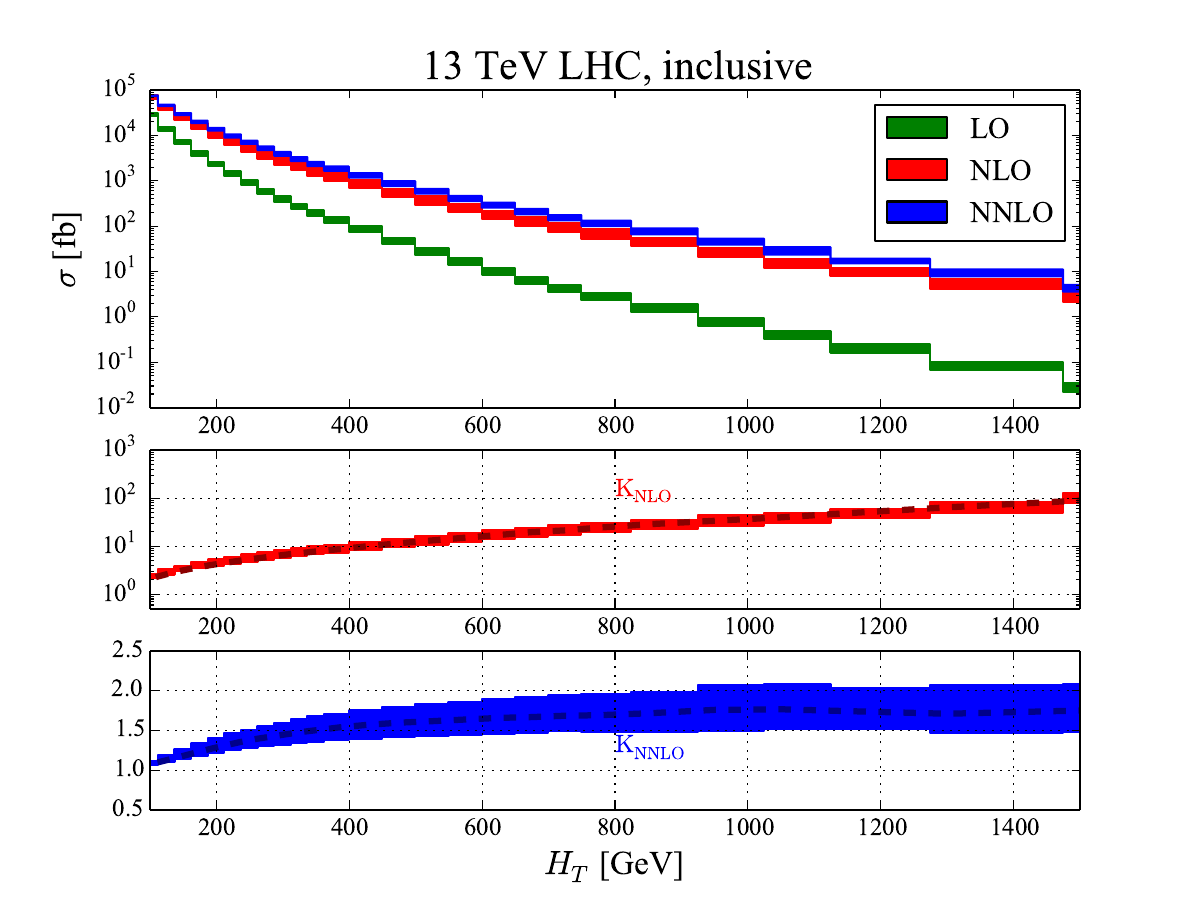} 
    \includegraphics[width=.49\linewidth]{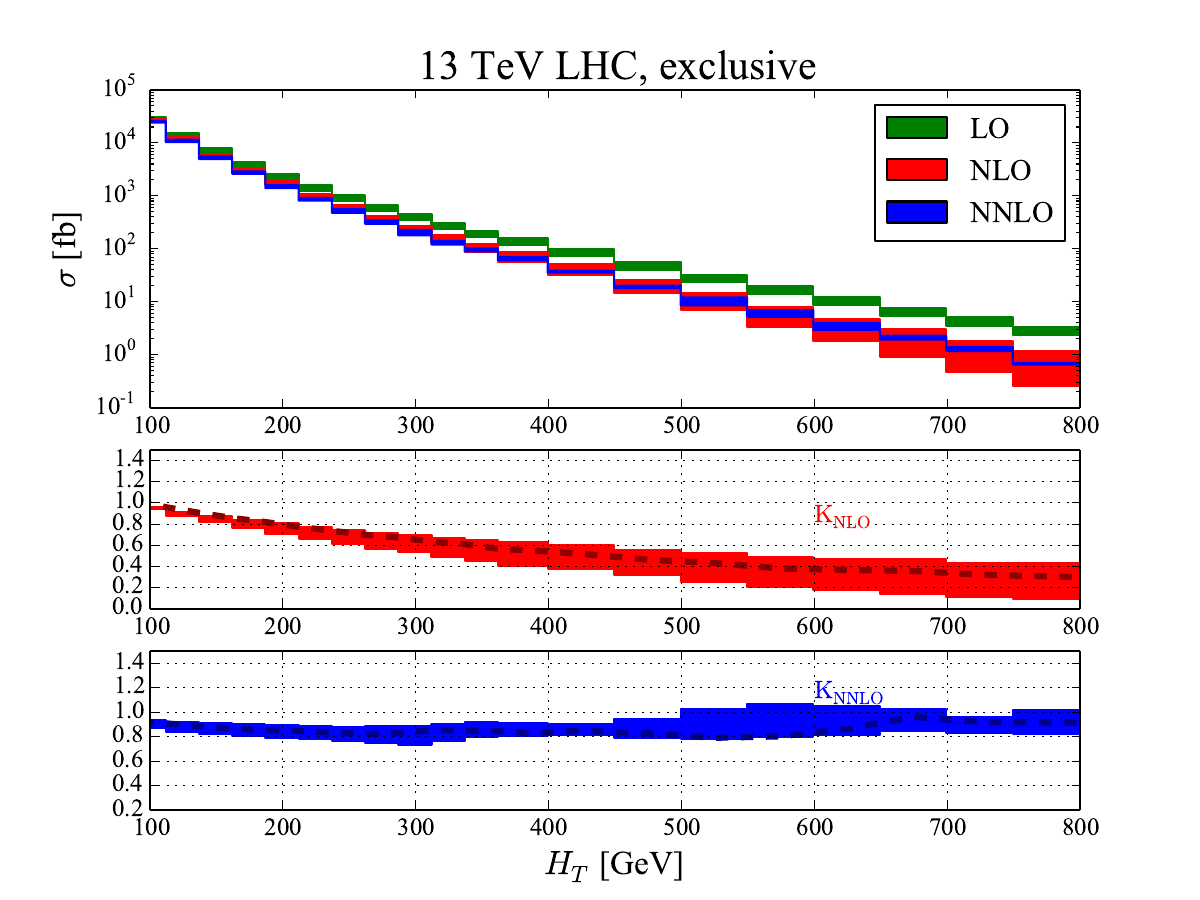} 
    \caption{Plots of the $H_T$ distribution for the following scenarios: 8 TeV inclusive 1-jet bin (upper left), 8 TeV exclusive 1-jet bin (upper right), 13 TeV inclusive 1-jet bin (lower left), 13 TeV exclusive 1-jet bin (lower right).  In each plot the upper inset shows the LO, NLO and NNLO distributions, while the lower inset shows $K_{\NLO}$ and $K_{\NNLO}$.  The bands indicate the scale variation, while the dashed lines in the lower panel indicate the result for the central scale choice.}
    \label{fig:HT}
\end{figure}

We next consider the distribution of $H_T$, which we define as the scalar sum of the transverse momenta of all jets that pass the acceptance cuts of Eq.~(\ref{eq:cuts1}).  The LO, NLO and NNLO results are shown for both collision energies, and for both the inclusive and exclusive 1-jet bins, in Fig.~\ref{fig:HT}.  The giant $K$-factor at NLO is clearly visible in the inclusive case for both collision energies.  The corrections grow to a factor of 100 for $H_T=1.2$ TeV at 8 TeV and for $H_T=1.5$ TeV at 13 TeV.  The NLO scale uncertainties are approximately $\pm 30\%$ for $H_T>1$ TeV for both 8 and 13 TeV collisions.  Although still large, the NNLO corrections are much smaller than the observed corrections at NLO.  For both 8 TeV and 13 TeV collisions the corrections grow to $+75\%$ for the central scale choice when $H_T> 1$ TeV, with a residual scale dependence at the $\pm 15\%$ level.   Further theoretical work is needed to bring the residual error from uncalculated QCD corrections down to the percent level for the $H_T$ distribution.  It would be interesting to compare the NNLO corrections obtained here with a merged $W$+jets sample, for which the perturbative expansion of the $H_T$ distribution is more tame~\cite{Hoeche:2012yf,NLOQCDEW}.  We note that the $H_T$ distribution in the exclusive one-jet bin is identical to the $p_T^{J_1}$ distribution in the exclusive one-jet bin, since there is only one reconstructed jet in this case.
 
\begin{figure}[h]
    \includegraphics[width=.49\linewidth]{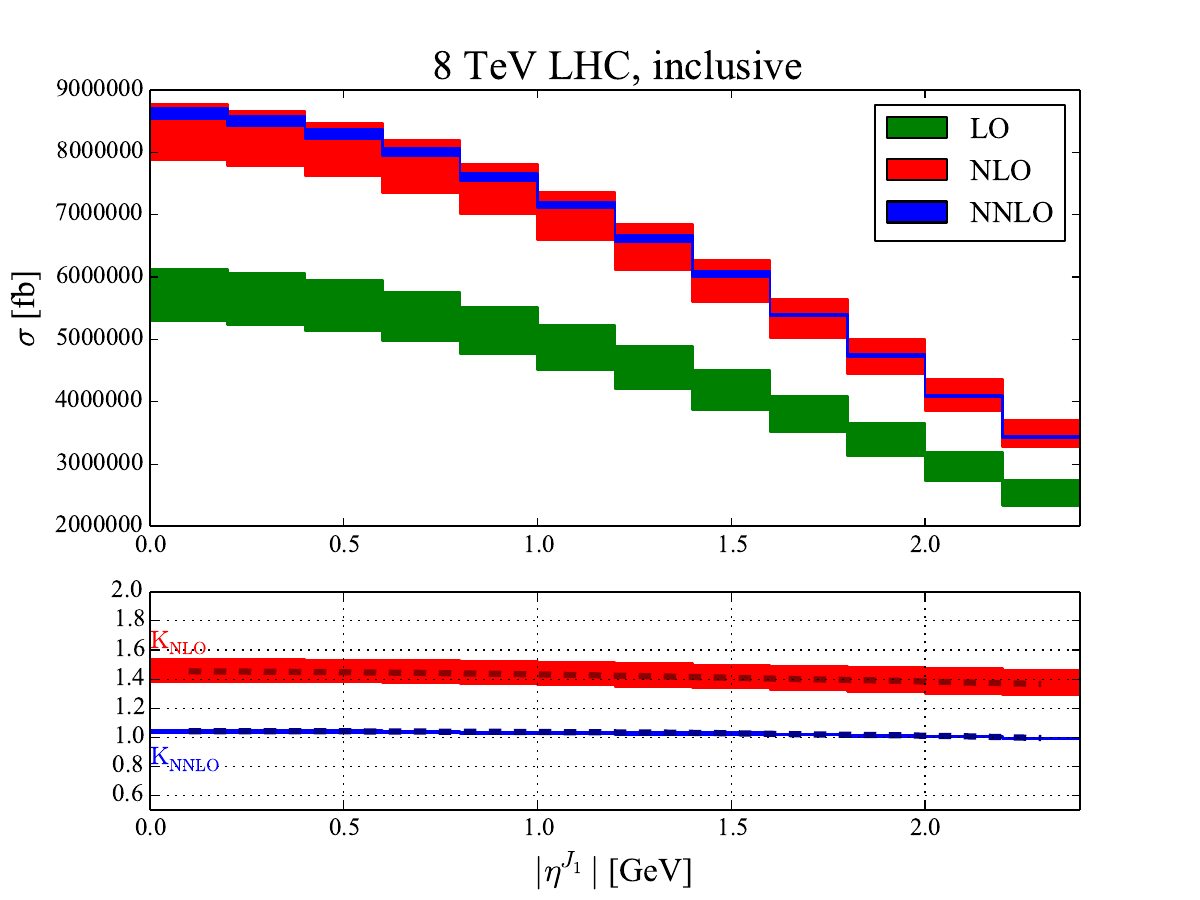} 
    \includegraphics[width=.49\linewidth]{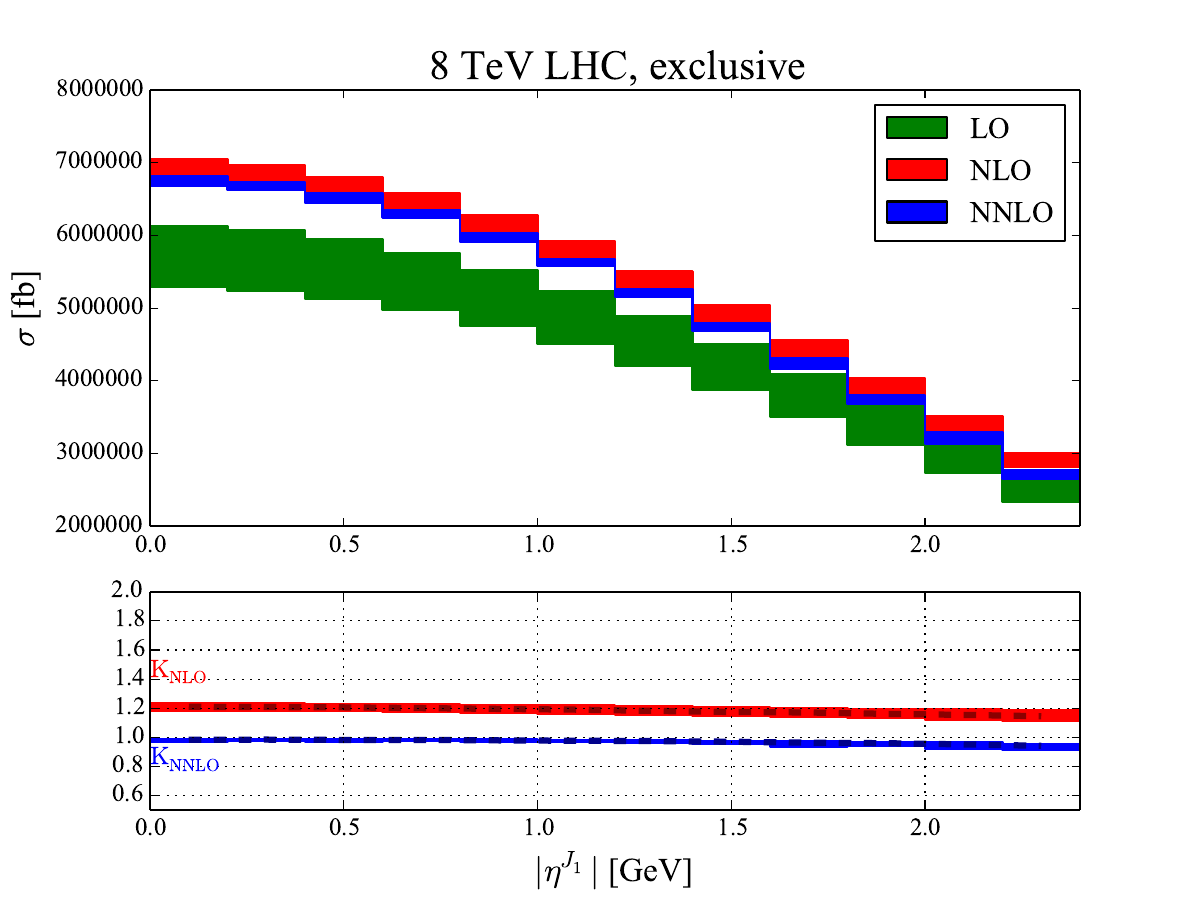} 
    \includegraphics[width=.49\linewidth]{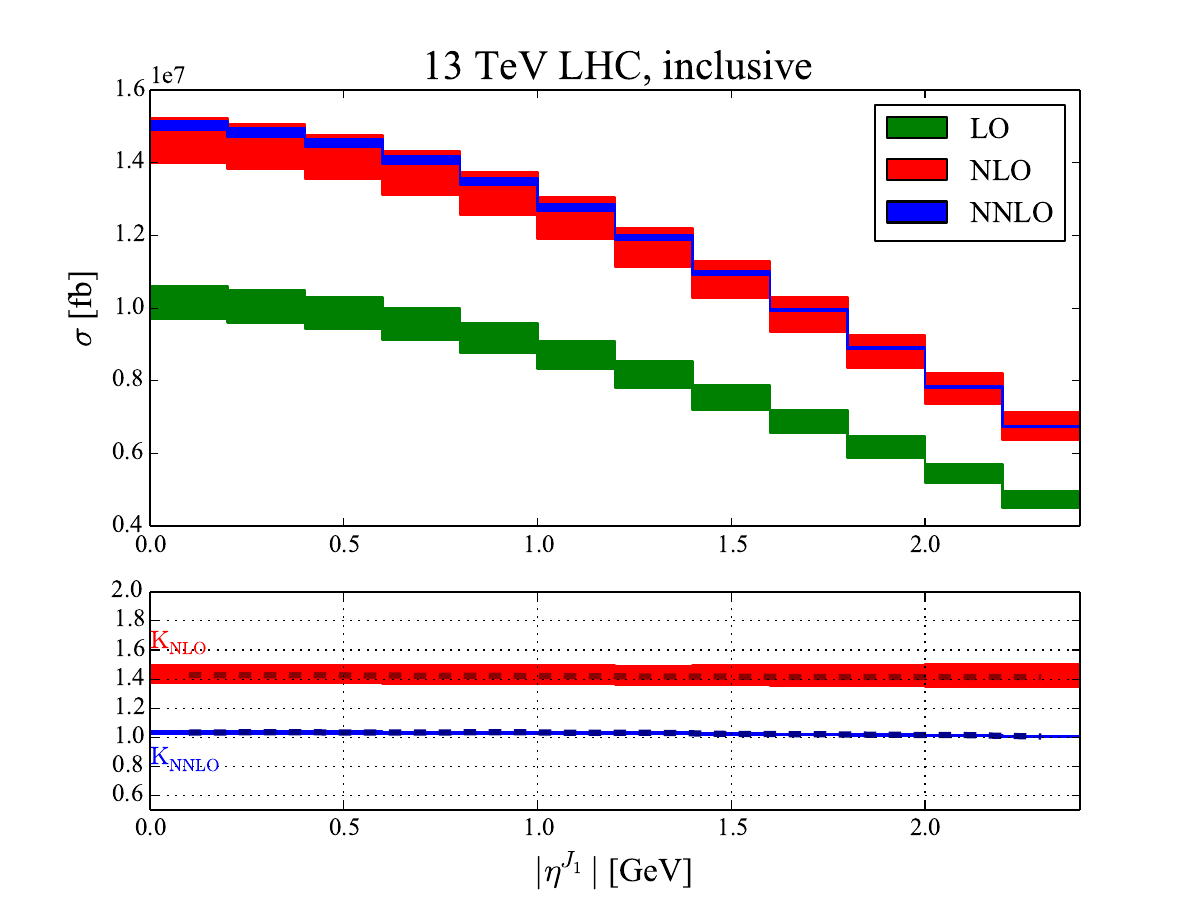} 
    \includegraphics[width=.49\linewidth]{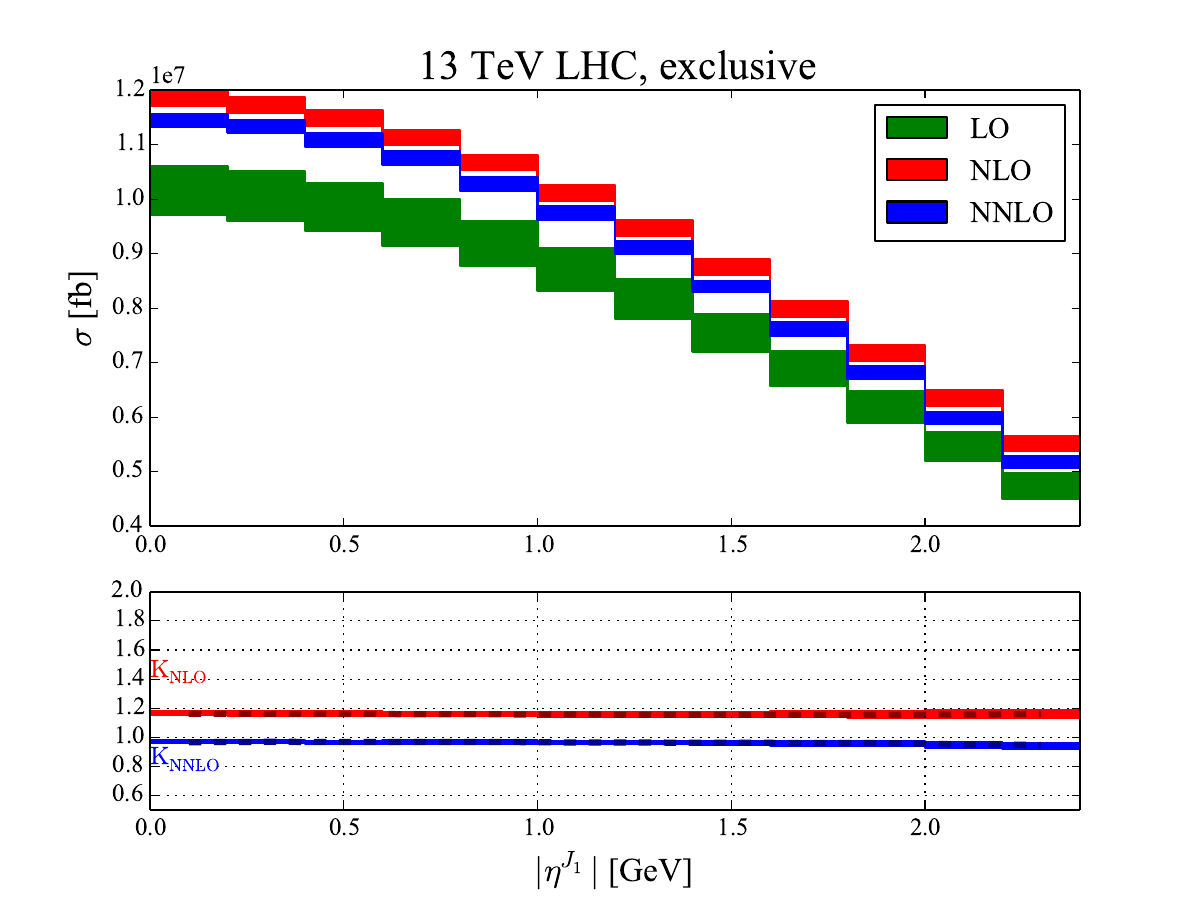} 
    \caption{Plots of the leading-jet pseudorapidity distribution for the following scenarios: 8 TeV inclusive 1-jet bin (upper left), 8 TeV exclusive 1-jet bin (upper right), 13 TeV inclusive 1-jet bin (lower left), 13 TeV exclusive 1-jet bin (lower right).  In each plot the upper inset shows the LO, NLO and NNLO distributions, while the lower inset shows $K_{\NLO}$ and $K_{\NNLO}$.  The bands indicate the scale variation, while the dashed lines in the lower panel indicate the result for the central scale choice.}
    \label{fig:etaJ1}
\end{figure}

We next study the pseudorapidity of the leading jet in Fig.~\ref{fig:etaJ1}.  At NLO the corrections are flat as a function of pseudorapidity, increasing the LO result by a factor of about 1.4 for both 8 TeV and 13 TeV collisions.  The NNLO corrections further increase the result by a few percent, following the pattern of corrections observed for the fiducial cross section. The estimated theoretical uncertainties decrease from  $\pm 5\%$ at NLO to $\pm 1\%$ at NNLO for both collision energies. 
The corrections in the exclusive 1-jet bin have a similar shape as the corrections in the inclusive 1-jet bin, but a different magnitude.  The NNLO correction reduces the result by a few percent, the same result observed for the fiducial cross section.
 
\begin{figure}[h]
    \includegraphics[width=.49\linewidth]{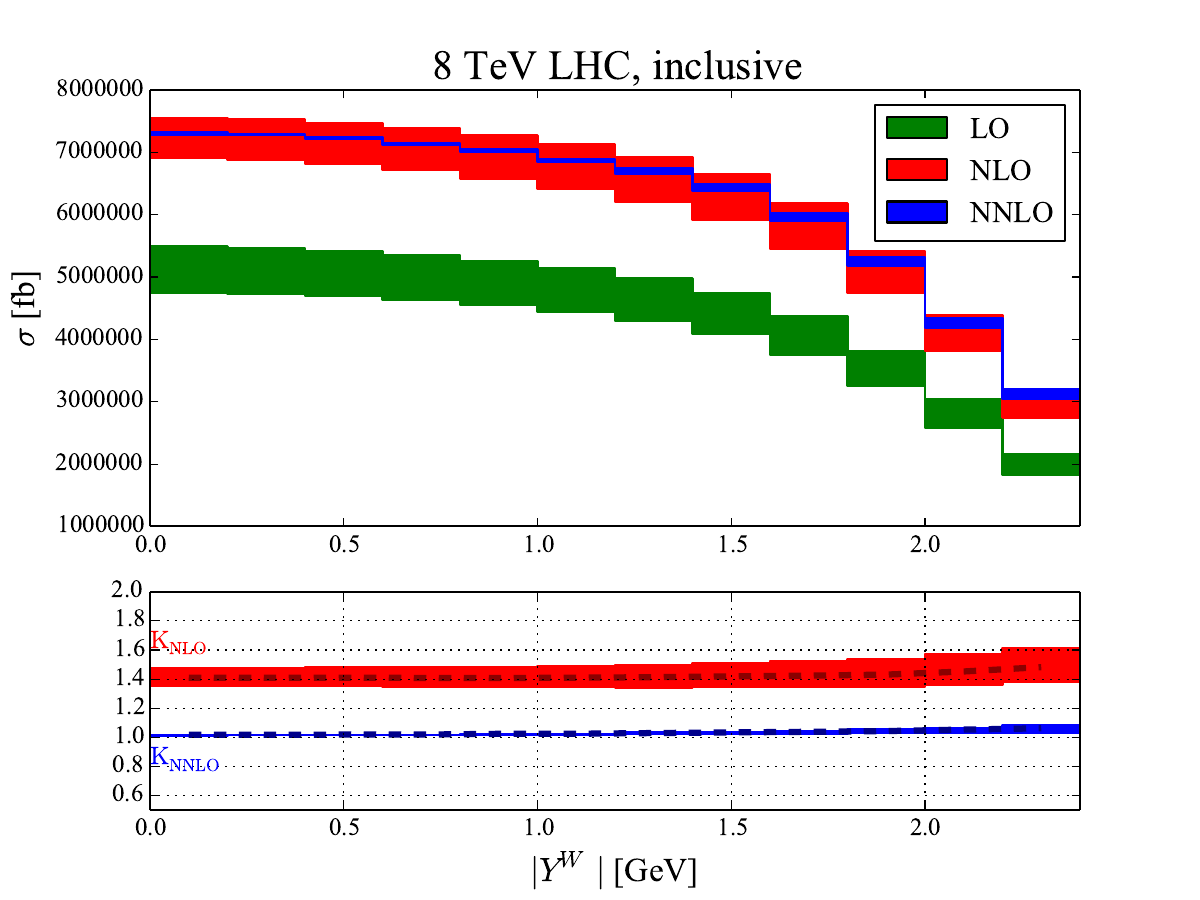} 
    \includegraphics[width=.49\linewidth]{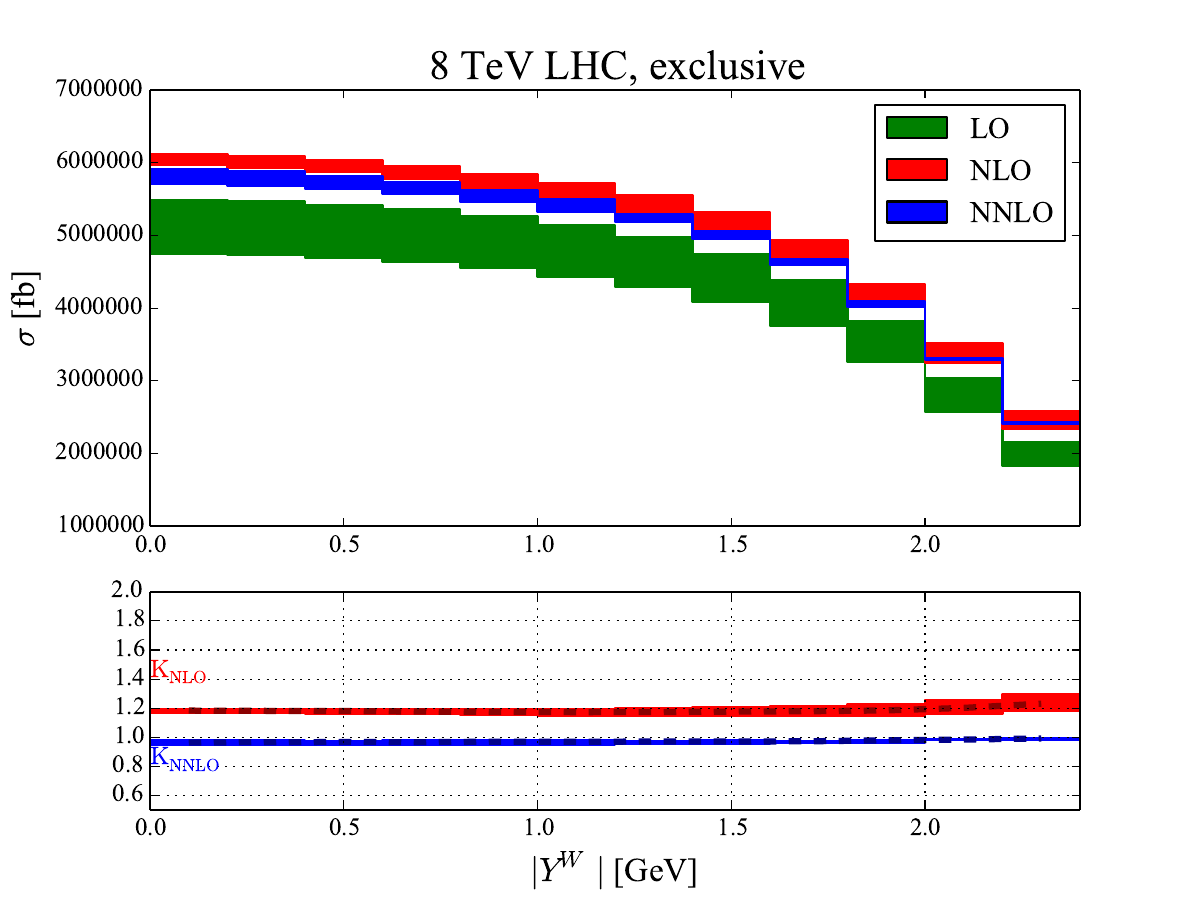} 
    \includegraphics[width=.49\linewidth]{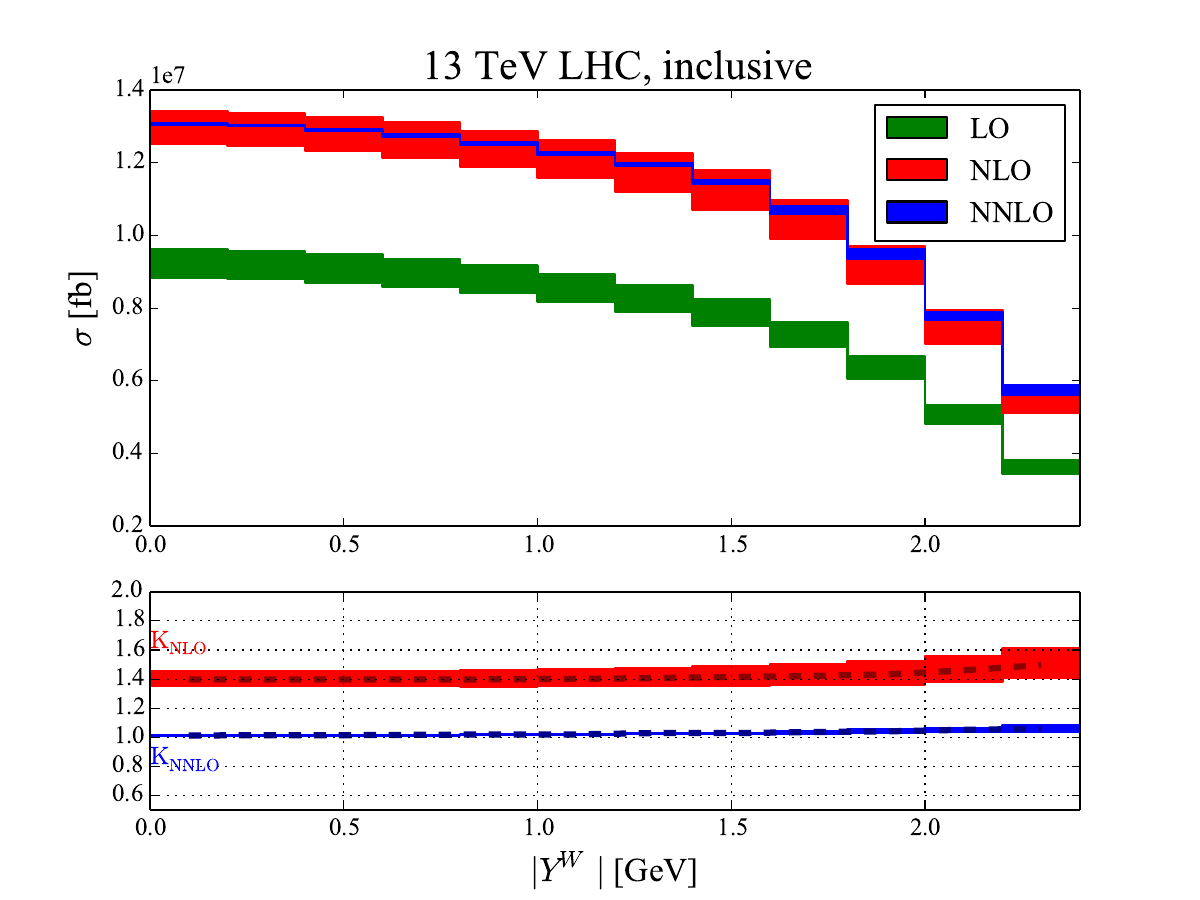} 
    \includegraphics[width=.49\linewidth]{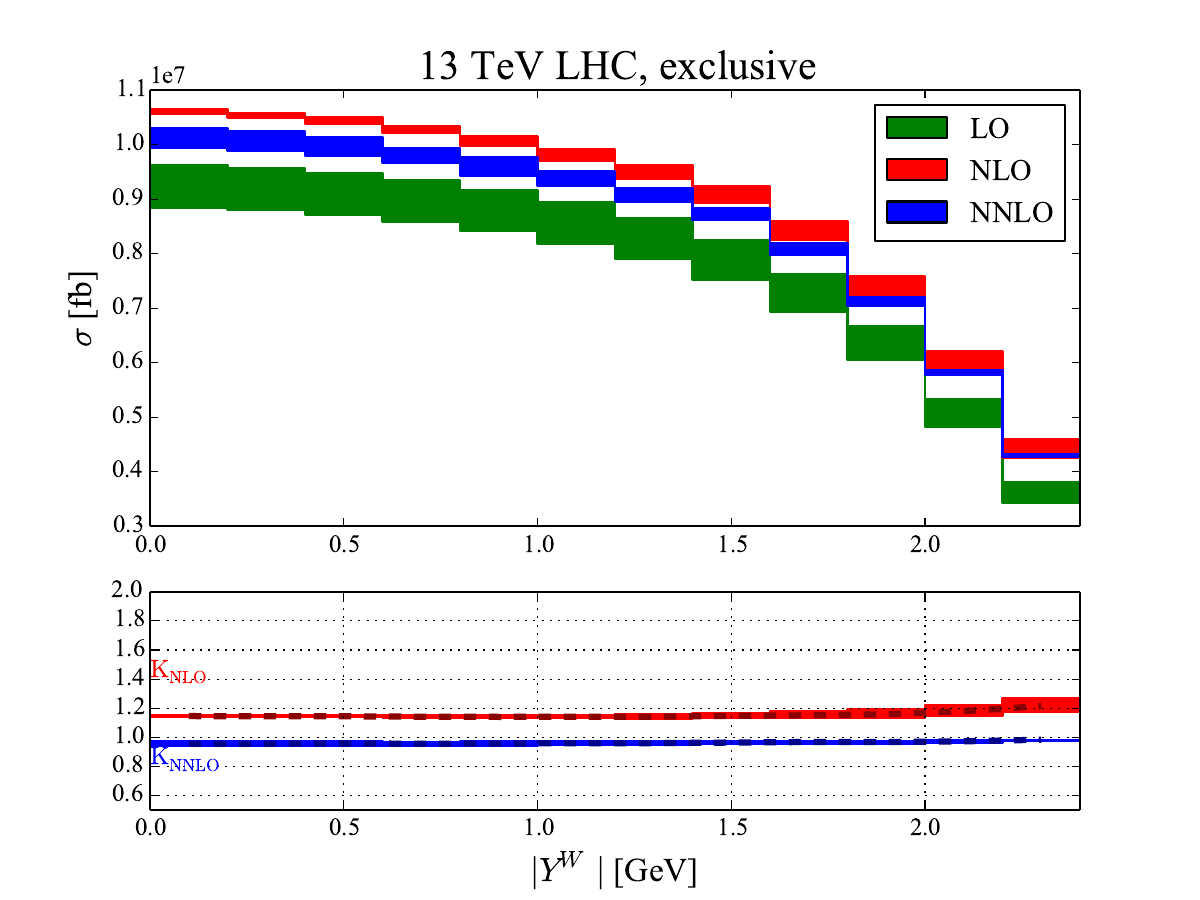} 
    \caption{Plots of the $W$-boson rapidity distribution for the following scenarios: 8 TeV inclusive 1-jet bin (upper left), 8 TeV exclusive 1-jet bin (upper right), 13 TeV inclusive 1-jet bin (lower left), 13 TeV exclusive 1-jet bin (lower right).  In each plot the upper inset shows the LO, NLO and NNLO distributions, while the lower inset shows $K_{\NLO}$ and $K_{\NNLO}$.  The bands indicate the scale variation, while the dashed lines in the lower panel indicate the result for the central scale choice.}
    \label{fig:Wrap}
\end{figure}

We next study the rapidity distribution of the $W$-boson in Fig.~\ref{fig:Wrap}. Both the NLO and NNLO corrections are remarkably flat as a function of rapidity.  For both 8 and 13 TeV collisions, a slight increase of the NLO $K$-factor from 1.4 to 1.5 is found as the rapidity is increased from $Y^W=0$ to $|Y^W|=2.5$.  For both collision energies the NNLO correction increases the cross section by a few percent. The NLO scale variation is approximately $\pm  5\%$ independent of rapidity, while the remaining NNLO scale uncertainty shows the same pattern as the fiducial cross section in Table~\ref{tab:fidinc}.  As for the jet pseudorapidity distribution, the corrections in the exclusive 1-jet bin have a similar shape as the corrections in the inclusive 1-jet bin, but a different magnitude.  The NNLO correction decreases the result for central rapidities by about 3\%.  There is a slight increase in the NLO and NNLO $K$-factors as $|Y_W|$ is increased.
 
\begin{figure}[h]
    \includegraphics[width=.49\linewidth]{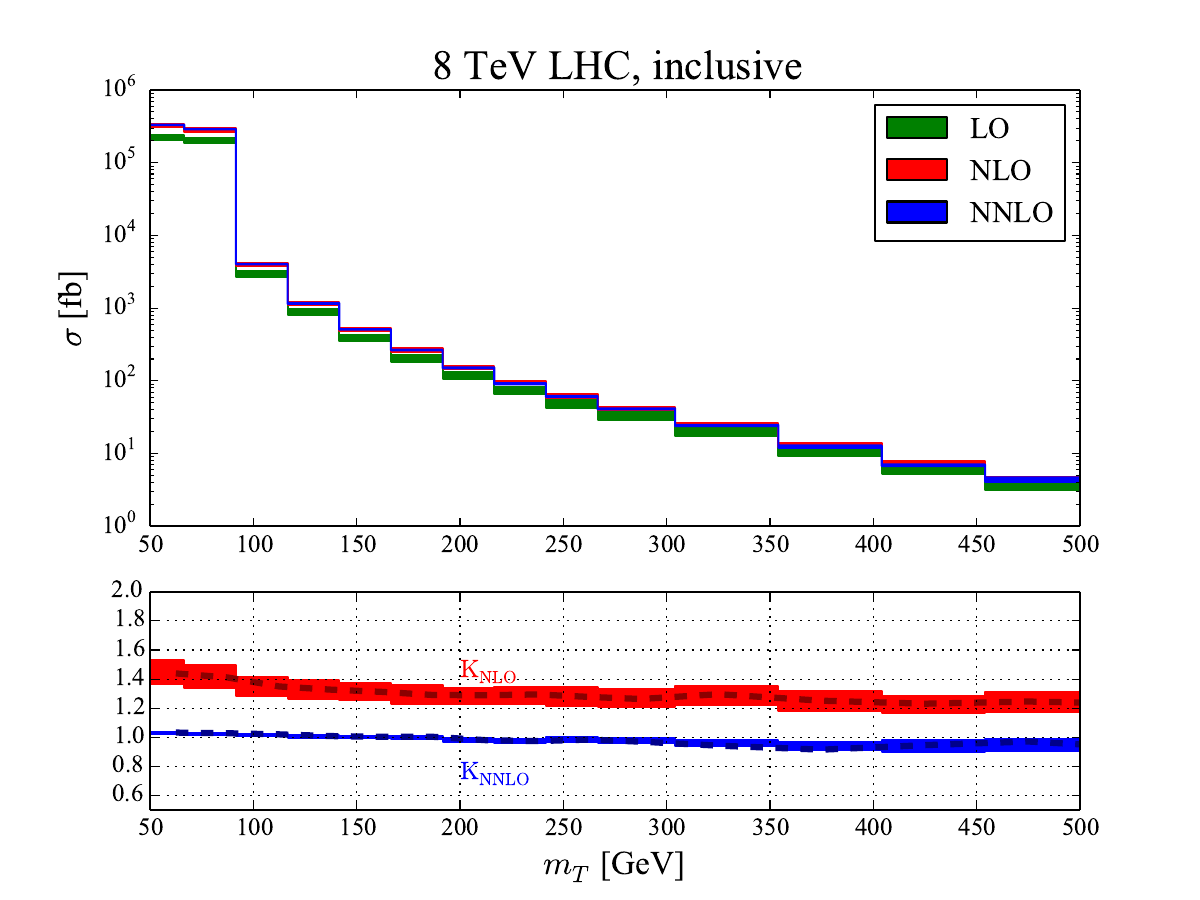} 
    \includegraphics[width=.49\linewidth]{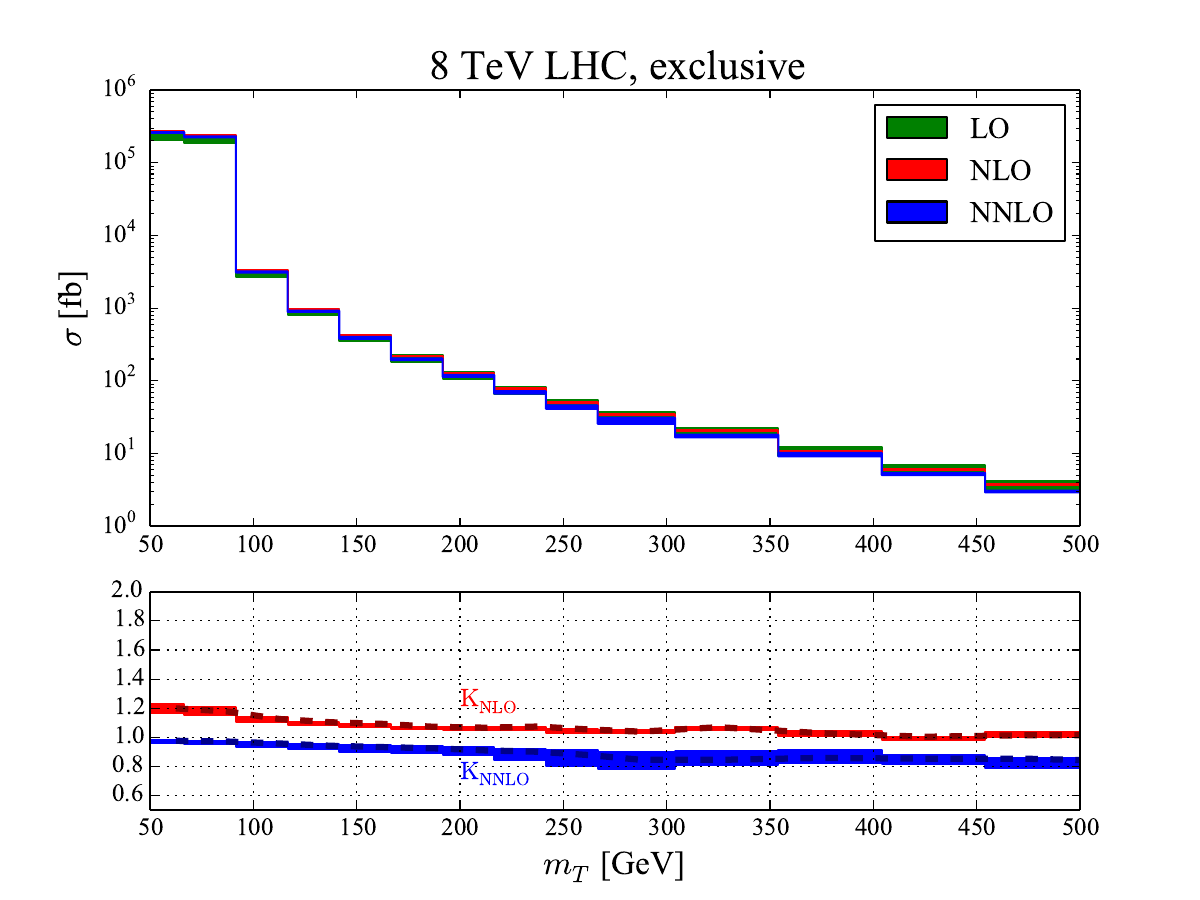} 
    \includegraphics[width=.49\linewidth]{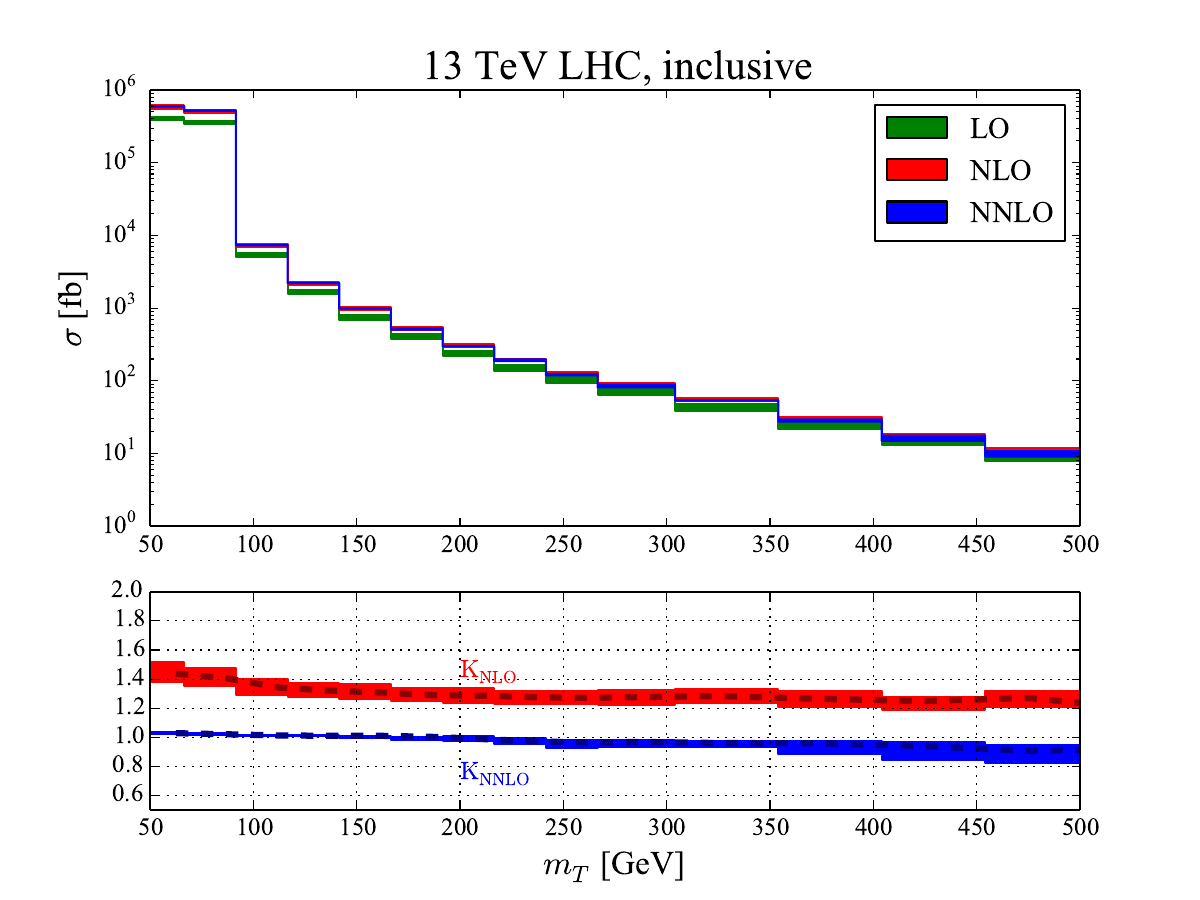} 
    \includegraphics[width=.49\linewidth]{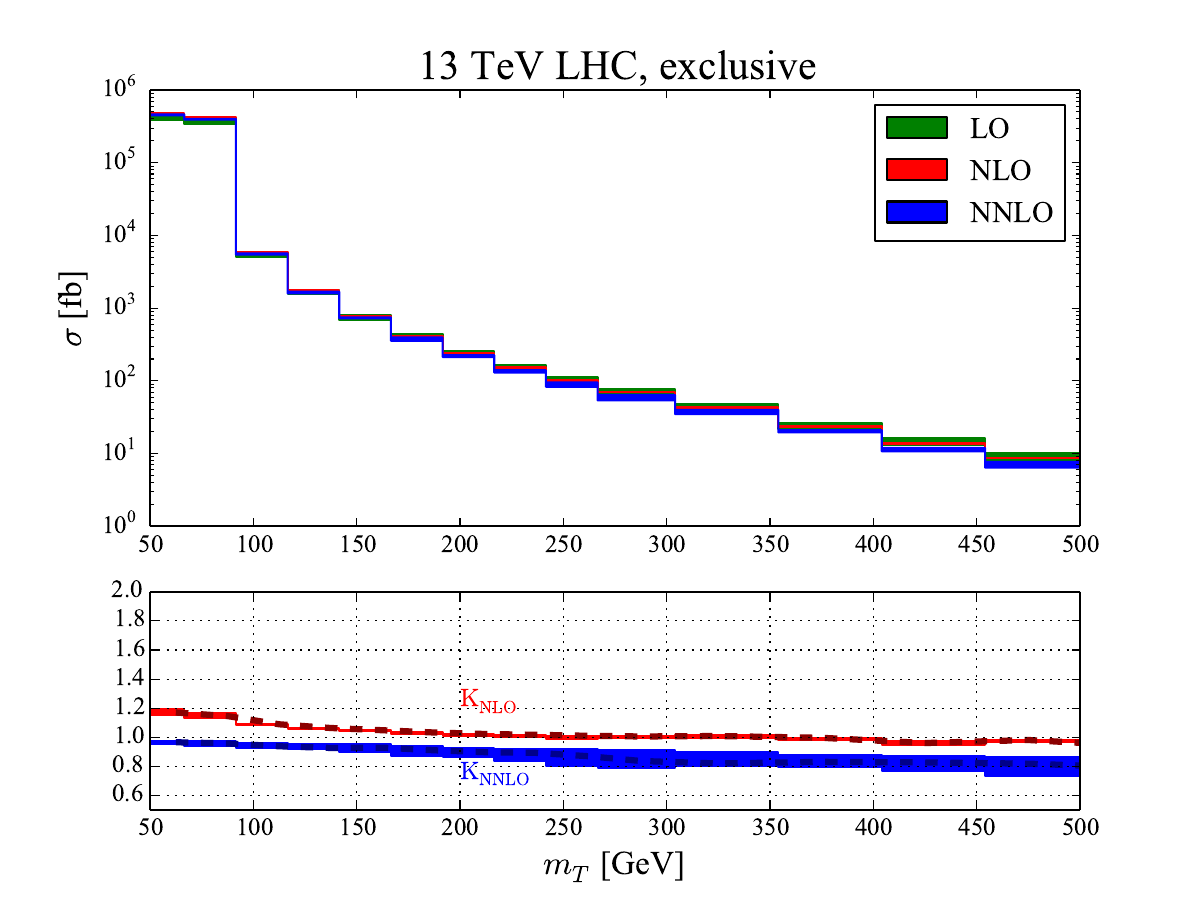} 
    \caption{Plots of the transverse mass distribution for the following scenarios: 8 TeV inclusive 1-jet bin (upper left), 8 TeV exclusive 1-jet bin (upper right), 13 TeV inclusive 1-jet bin (lower left), 13 TeV exclusive 1-jet bin (lower right).  In each plot the upper inset shows the LO, NLO and NNLO distributions, while the lower inset shows $K_{\NLO}$ and $K_{\NNLO}$.  The bands indicate the scale variation, while the dashed lines in the lower panel indicate the result for the central scale choice.}
    \label{fig:mT}
\end{figure}

Finally, we show in Fig.~\ref{fig:mT} the results for the transverse mass distribution.  To explain the observed distributions, we recall that the $m_T$ distribution for on-shell $W$-boson production exhibits a Jacobian peak at $M_W$.  Higher values of $m_T$ beyond the $W$-boson mass are generated by non-zero $p_T^W$, as well as the $W$-boson width.  However, the high $m_T$ region requires a very large $p_T^W$, as is well known~\cite{Baur:2003jy}, leading to the strong peak of the distribution around $M_W$.  The NLO correction reaches approximately 40\% for $m_T$ near the lower boundary of 50 GeV for both collision energies. The NNLO correction is much smaller, reaching a few percent for both 8 TeV and 13 TeV collisions.  The NNLO correction decreases slightly as $m_T$ increases.  The scale variation at the peak of the $m_T$ distribution decreases from $\pm 5\%$ at NLO to approximately $\pm 1\%$ at NNLO.  The $m_T$ distribution and higher-order corrections for the exclusive one-jet bin follow the same pattern as seen  for the inclusive one-jet bin.

\section{Numerical results for the boosted selection}
\label{sec:numboost}

We next study the cross section assuming the selection cuts of Eq.~(\ref{eq:cuts2}) for 8 TeV LHC collisions.  For this set of cuts the leading jet is required to be boosted: $p_T^{J_1}>500$ GeV.  This is a region of interest in studies of jet substructure~\cite{boostcms,boostatlas}. Two distinct categories of events pass these selection cuts: events where the leading jet balances in the transverse plane against a high-$p_T$ $W$-boson, and events with back-to-back jets together with the emission of a soft and/or collinear $W$-boson.  The first type of event occurs at leading order in the perturbative expansion for the $W$+jet process, while the second type of event first occurs at NLO.  We are interested here in the radiation pattern of the $W$-boson and the charged lepton that comes from its decay when these two event categories are combined.  How often are the lepton and $W$-boson emitted along the direction of a jet, and how stable are these predictions with respect to QCD corrections?
 
We begin by showing in Table~\ref{tab:fidboost} the fiducial cross sections assuming the cuts of Eq.~(\ref{eq:cuts2}).  The correction when going from LO to NLO is large, reaching 280\%, due to the new event category that appears at NLO.  The NNLO correction is smaller, and increases the NLO result by 16\%.  The scale variation decreases from approximately $\pm 20\%$ at NLO to the asymmetric range $(+3\%,-7\%)$ at NNLO.  We note that the NNLO correction is contained within the NLO scale variation band, indicating convergence of the perturbative expansion.

\begin{table}[h]
\begin{tabular}{|c|c|c|c|c|c|}
\hline
 & $\sigma_{\LO}$ (fb) & $\sigma_{\NLO}$ (fb) & $\sigma_{\NNLO}$ (fb) & $K_{\NLO}$ & $K_{\NNLO}$ \\
 \hline\hline
8 TeV & $56.53^{+13.25}_{-10.04}$ & $160.4^{+34.5}_{-26.5}$ & $186.7^{+5.4}_{-11.9}$ & 2.84 & 1.16 \\
\hline
\end{tabular}
\caption{Fiducial cross sections for the boosted cuts of Eq.~(\ref{eq:cuts2}) for 8 TeV LHC collisions.  The scale errors are shown for the LO, NLO and NNLO cross sections.  The $K$-factors are shown for the central scale choice.}
\label{tab:fidboost}
\end{table} 

\begin{figure}[h]
    \includegraphics[width=.75\linewidth]{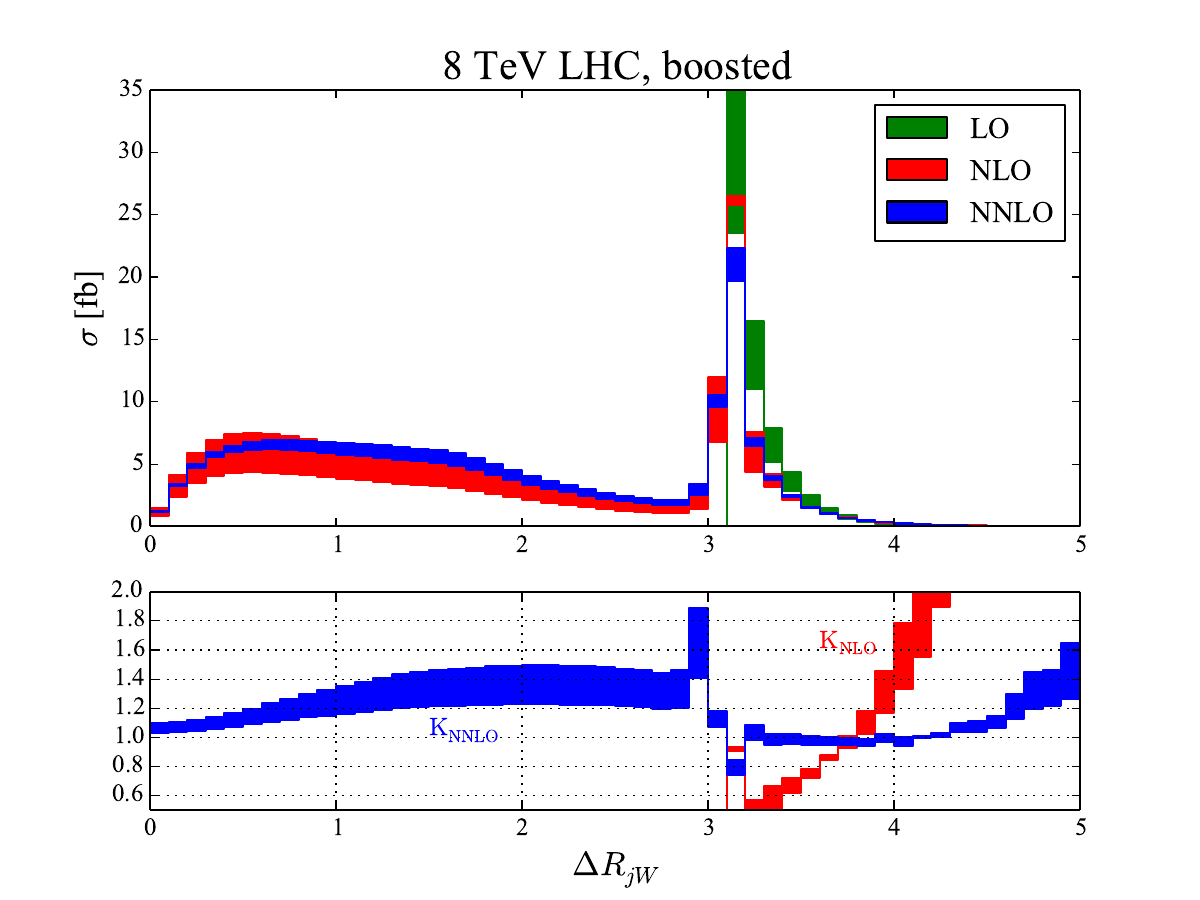} 
    \caption{Plot of the $\Delta R_{jW}$ distribution for 8 TeV LHC collisions.   The upper inset shows the LO, NLO and NNLO distributions, while the lower inset shows $K_{\NLO}$ and $K_{\NNLO}$.  The bands indicate the scale variation.}
    \label{fig:delrjW}
\end{figure}

Next we study the separation between the charged lepton and the closest jet, as well as the $W$-boson and the closest jet.  The distances are measured using $\Delta R_{jl}$ and $\Delta R_{jW}$ respectively, where $\Delta R_{xy}$ is defined in Eq.~(\ref{eq:delr}).  The $\Delta R_{jW}$ distribution is shown in Fig.~(\ref{fig:delrjW}).  At leading order the $W$-boson and jet must be back-to-back in the transverse plane, leading to the kinematic requirement $\Delta R_{jW} \geq \pi$.  This restriction is relaxed at NLO.  While the distribution of events in the $\Delta R_{jW}< \pi$ region is slightly peaked toward lower $\Delta R_{jW}$, it is still fairly broad, indicating that there is no strong enhancement for $W$-bosons emitted collinear to a jet.  In the region $\Delta R_{jW}<\pi$ the NNLO correction varies from $+10\%$ to $+35\%$ as $\Delta R_{jW}$ is increased.  The perturbative expansion rapidly changes as the point $\Delta R_{jW} = \pi$ is crossed, with the NNLO correction changing from $+60\%$ directly below this value to $-20\%$ directly above.  This region is sensitive to soft gluon effects, as it represents the leading-order kinematic boundary.  Fixed-order perturbation theory contains a large logarithm of the $\Delta R_{jW}$ bin size, and diverges as the bin size is taken to zero.  Above this region the NNLO corrections remain at the percent level until the upper phase-space edge is reached.  The NNLO scale uncertainty is approximately $\pm 7-10\%$ in the region  $2 < \Delta R_{jW} < 3$, with the larger values occurring for the upper edge of this range.  For  $\Delta R_{jW} > \pi$, and away from the LO kinematic boundary, the NNLO scale uncertainty is in the $\pm 3-4\%$ range, until it increases near the upper boundary of phase space.

Another interesting quantity to consider is the fraction of events with $\Delta R_{jW} < 3$.  This region is separated enough from the LO kinematic boundary to be computed in fixed-order perturbation theory.  Defining the fraction $F_{jW} = \sigma(\Delta R_{jW}<3) / \sigma_{total}$, we find 
\begin{equation}
F_{jW}^{\NLO} = 0.67^{+0.06}_{-0.05}, \;\;\; F_{jW}^{\NNLO} = 0.74^{+0.02}_{-0.02},
\end{equation}
where the superscripts and subscripts indicate the scale-variation error.  The majority of events do not feature a back-to-back $W$-boson and jet, but rather contain predominantly a di-jet configuration with the emission of a softer $W$-boson.  This fraction is stable with respect to QCD radiative corrections.
 
\begin{figure}[h]
    \includegraphics[width=.75\linewidth]{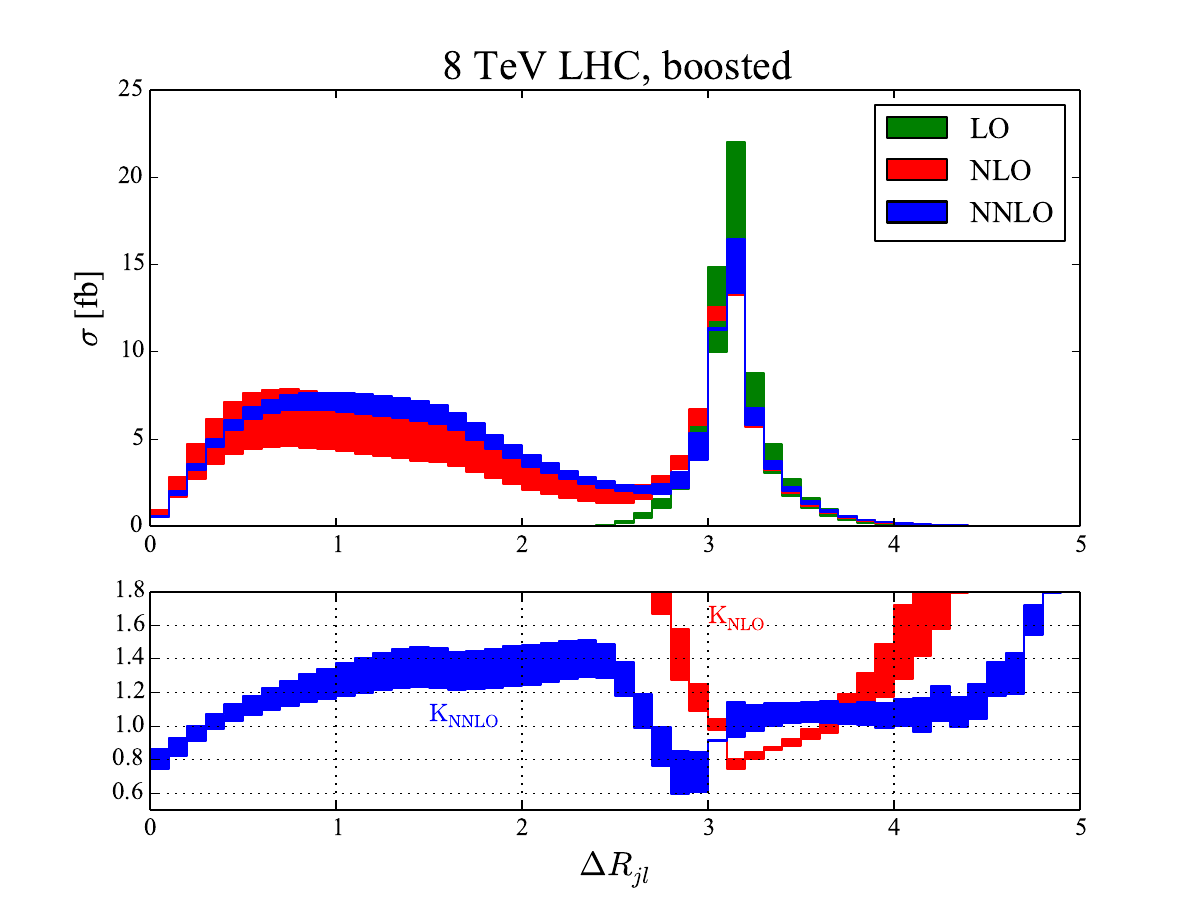} 
    \caption{Plot of the $\Delta R_{jl}$ distribution for 8 TeV LHC collisions.   The upper inset shows the LO, NLO and NNLO distributions, while the lower inset shows $K_{\NLO}$ and $K_{\NNLO}$.  The bands indicate the scale variation.}
    \label{fig:delrjl}
\end{figure}

We also show in Fig.~\ref{fig:delrjl} the $\Delta R$ distribution between the lepton and the closest jet.  The pattern of corrections for this observable is very similar to the result seen for $\Delta R_{jW}$.  Although the LO distribution does not  vanish below $\Delta R_{jW}=\pi$ since the lepton is not emitted exactly along the $W$-boson correction, there are still extremely large corrections below this boundary.  In the region $\Delta R_{jl}<\pi$ the NNLO correction varies from $+10\%$ to $+45\%$ as $\Delta R_{jl}$ is increased. The QCD perturbative expansion again shows sensitivity to soft gluon effects near this region.  Since at LO the $W$-boson is highly boosted with $p_T^W>500$ GeV, the lepton is emitted preferentially along the $W$-boson direction, and the $\Delta R_{jl} \approx \pi$ region remains sensitive to the kinematic boundary appearing for $\Delta R_{jW}=\pi$.  Above this boundary the NNLO corrections vary from $7-10\%$, increasing further near the upper kinematic limit where the cross section vanishes.  The theoretical uncertainty as estimated by scale variation ranges from 5\% to 20\% for $\Delta R_{jl} \approx \pi$, and is at the 15\% level above.  We again compute the fraction of events with $\Delta R_{jW} < 3$, finding
\begin{equation}
F_{jl}^{\NLO} = 0.74^{+0.05}_{-0.04}, \;\;\; F_{jl}^{\NNLO} = 0.77^{+0.02}_{-0.02}.
\end{equation}
The values and stability in perturbation theory are similar to those found for $F_{jW}$.

\section{Summary and conclusions} 
\label{sec:conc} 
 
In this manuscript we have performed a detailed phenomenological study of the $W$+jet process through NNLO in perturbative QCD.   We have presented results for 8 TeV and 13 TeV LHC collisions, and for multiple selection criteria: a set of general acceptance cuts and one focusing on the high transverse momentum jet region.  Both the inclusive and exclusive one-jet bins have been considered.  For most observables studied and in most phase space regions, the QCD perturbative expansion is under good theoretical control.  Away from kinematic boundaries the NNLO corrections in the inclusive 1-jet bin are 5\% in the bulk of phase space, reaching 15\% in the high-$p_T$ tail of the leading jet transverse momentum region.  The only exception is the $H_T$ distribution, for which corrections reaching 75\% in the TeV energy range are still observed.  It would be interesting to compare the NNLO predictions with merged samples of $W$+jets at NLO.  Our NNLO calculation for the inclusive one-jet bin contains events with both two and three jets.  However, the NLO corrections to the $W$+multi-jet process are known through $W$+5 jets~\cite{Bern:2013gka}.  It is therefore possible to include more real-emission contributions to the $H_T$ distribution using a merged sample of $W$+multi-jet events at NLO.

The corrections in the exclusive one-jet bin contain large jet-veto logarithms in the high energy region, and consequently the QCD perturbative expansion is under poorer control.  At NLO, the cross section decreases by up to an order of magnitues in the high-$p_T$ region.  The NNLO corrections improve the situation, and the corrections are much smaller.  However, it would be useful to compare these predictions with those obtained from resummation-improved perturbation theory.  For the boosted selection, a new phase space region opens up in the $\Delta R$ distributions at NLO, leading to large corrections at this order.  The perturbative expansion stabilizes when the NNLO corrections are included.  There is no strong peak for collinear emission of the $W$-boson along the jet directioK\"uhnn.

We believe that QCD corrections to the $W$+jet process are now under good theoretical control.  We have indicated several possible future extensions of our results.  Analyses of the $W$+jet process in 8 TeV collisions are ongoing, and studies at 13 TeV have begun.  Given the importance of this process for the LHC program, we hope there is continued theoretical investigation using the results presented here as a basis for future work.
 
\section*{Acknowledgments}
We thank J.~Huston and T.~LeCompte for helpful communications.  R.~B. is supported by the DOE contract DE-AC02-06CH11357.  X.~L. is supported by the DOE grant DE-FG02-93ER-40762.  F.~P. is supported by the DOE grants DE-FG02-91ER40684 and DE-AC02-06CH11357.  This research used resources of the National Energy Research Scientific Computing Center, a DOE Office of Science User Facility supported by the Office of Science of the U.S. Department of Energy under Contract No. DE-AC02-05CH11231.  It also used resources of the Argonne Leadership Computing Facility, which is a DOE Office of Science User Facility supported under Contract DE-AC02-06CH11357.

\end{document}